\documentclass[referee,a4paper,12pt,structabstract]{swsc} 

\usepackage{graphicx}
\usepackage{txfonts}
\usepackage{subfigure}
\usepackage{epstopdf}
\usepackage[authoryear,round]{natbib}
\usepackage[backref]{hyperref}
\usepackage{url}

\hypersetup{colorlinks=true,citecolor=cyan,urlcolor=cyan,linkcolor=blue}

\begin{document}


   \title{A new technique for observationally derived boundary conditions for space weather}
   
   \titlerunning{Boundary conditions for space weather}

   \authorrunning{Pagano, Mackay, Yeates}

   \author{Paolo Pagano\inst{1}
          \and
          Duncan Hendry Mackay\inst{1}
          \and
          Anthony Robinson Yeates\inst{2}
          }

   \institute{School of Mathematics and Statistics, University of St. Andrews, North Haugh, St. Andrews, Fife, Scotland KY16 9SS, UK\\
              \email{pp25@st-andrews.ac.uk}
         \and
             Department of Mathematical Sciences, Durham University, Durham, DH1 3LE, UK\\
             }


 
  \abstract
   {In recent years, space weather research has focused 
on developing modelling techniques to predict the arrival time and properties of coronal mass 
ejections (CMEs) at the Earth.}
   {The aim of this paper is to propose a new modelling technique suitable for the next generation of Space 
Weather predictive tools that is both efficient and accurate.
The aim of the new approach is to provide interplanetary space
weather forecasting models with accurate time dependent boundary conditions of erupting magnetic flux ropes
in the upper solar corona.}
   {To produce boundary conditions, we couple two different modelling techniques, MHD 
simulations and a quasi-static non-potential evolution model.
Both are applied on a spatial domain that covers the entire solar surface, although they extend over a different radial distance.
The non-potential model uses a time series of observed synoptic magnetograms to drive the 
non-potential quasi-static evolution of the coronal magnetic field. This allows us to follow the formation and 
loss of equilibrium of magnetic flux ropes.
Following this a MHD simulation captures the dynamic evolution of the erupting flux rope,
when it is ejected into interplanetary space.}
  {The present paper focuses on the MHD simulations that follow the ejection of magnetic flux ropes to
4$R_\odot$. We first propose a technique for specifying the pre-eruptive plasma properties 
in the corona. Next, time dependent MHD simulations describe the 
ejection of two magnetic flux ropes, that produce time dependent boundary conditions for the magnetic field and plasma at 4$R_{\odot}$
that in future may be applied to interplanetary space weather prediction models.}
  {In the present paper, we show that the dual use of quasi-static non-potential magnetic field simulations
and full time dependent MHD simulations can produce realistic inhomogeneous boundary conditions for space weather 
forecasting tools. Before a fully operational model can be produced there are a number of technical and scientific challenges that still need to be addressed.
Nevertheless, we illustrate that coupling quasi-static and MHD simulations in this way can significantly reduce the computational time
required to produce realistic space weather boundary conditions.}


   \maketitle

\section{Introduction}
The solar corona is a highly dynamic environment where magnetic and plasma structures are 
continually evolving. It is a key region of the solar atmosphere that couples the solar
photosphere with interplanetary space. This coupling results in the solar wind and many
other perturbations of the interplanetary space plasma, referred to as space weather. 
The origin of these events can be traced back to flux emergence and flows at the solar photosphere. 
The corona stores vast amounts of free magnetic energy and the release of this 
energy in the form of coronal mass ejections (CMEs) initiates the largest and most violent 
perturbations of the near Earth environment, referred to as 
space weather or extreme space weather \citep{Hapgood2011}.
This topic has drawn the attention of national and international institutions 
due to the practical consequences that severe space weather events can have on national infrastructures as well as on space 
operations and missions \citep{Schrijver2015}.

CMEs produce major disruptions to the ambient magnetic and plasma properties of interplanetary space,
as their occurrence signifies the violent ejection of both magnetic flux and plasma from the low solar
corona. Often magnetic flux ropes in the low corona are considered the main progenitors of CMEs,
where an initial period of stability of the flux rope is followed by a fast and sudden ejection outwards into interplanetary space.
This scenario is supported by a number of observations \citep{Howard2014,Chintzoglou2015,Cheng2011,LiZhang2013}.
Some models explain the flux rope formation and the subsequent 
ejection with photospheric flows \citep{MackayVanBallegooijen2006A, Xia2014}, others focus on 
magnetic flux emergence from underneath the photosphere \citep{ArchontisHood2012}, and finally others
rely on the onset of MHD instabilities \citep{TorokKliem2005,Zuccarello2015}.
It should be noted however, that it is not unusual to observe 
CMEs bearing no apparent connection with structures in the low corona \citep{DHuys2014, Ouyang2015}.
A statistical study carried out by \citet{Hutton2015} showed that not all CMEs can be directly associated to magnetic flux ropes
and \citet{Vourlidas2013} measured that at least 40\% of CMEs are certainly associated with magnetic flux ropes,
while only a minor percentage are certainly not associated with any magnetic flux ropes.
The above discussion illustrates that there is a strong relationship between CMEs and flux ropes.

While the CME initiation mechanism is still under debate,
it is widely accepted that, no matter the origin of the CME, they travel outwards due to their own
propulsion until about $\sim4 R_{\odot}$ \citep{Gopalswamy2000b}, beyond which they are dragged out by the 
solar wind. \cite{Sachdeva2015} claim that the aerodynamic drag only becomes the dominant force after $15 R_{\odot}$,
although it can play a role at shorter radial distances. It therefore seems reasonable to assume that CMEs are not 
significantly coupled to the solar wind until after $4R_\odot$. Nevertheless, below $4 R_{\odot}$  \citet{Gopalswamy2003b}
and also at larger distances, CMEs can suffer both deflection or acceleration.
This may be due to the background corona, the interplanetary 
magnetic field or interaction with other eruptive events.

To mitigate the effects of space weather it is key to predict the arrival time and properties of CMEs at the Earth's magnetosphere.
The development of these predicting tools is not trivial and while current attempts have made significant advances, 
it is commonly accepted that more precision and more accuracy are required for effective predictions.
The most recent development of the drag-based model (DBM) \citep{Vr¨nak2013, Zic2015} is used to predict the arrival time of CMEs
assuming that beyond $20 R_{\odot}$ the aerodynamic drag force dominates the dynamics.
Moreover, the model ElEvoHI \citep{Rollett2016} empowers the DBM with an improved geometrical fitting of the CME.
\citet{Toth2005} introduced the Space Weather Modelling Framework (SWMF), a series of highly complex computer 
simulations that account for significantly different physical regimes and to this end utilises a number of communicating codes
\citep{Toth2012}.
The flexibility of the SWMF allows for the coupling of specific models with different codes to develop new modelling techniques,
as been done by \citet{Jin2017}.
\citet{Merkin2016} used a two different MHD codes to model the corona and the interplanetary space
and a common spherical shell doma at $20$ $R_{\odot}$ is used to couple the two codes.
Finally, the  3D MHD WSA-ENLIL model describes how the solar wind and interplanetary plasma act as a background for a 
kinematically inserted CME \citep{Odstrcil1996,OdstrcilPizzo1999a,OdstrcilPizzo1999b,Odstrcil2003,Odstrcil2004}.
In particular, the so called CME Cone Model  \citep{Xie2004, Xue2005, Michalek2006} is widely used in WSA-ENLIL to model the insertion of a CME 
into the solar wind. Its parameters can be tuned by coronagraph images \citep{Millward2013}, 
or data collected in situ at Mercury \citep{Baker2013}. 
Recently it has also been extended to describe Halo CMEs \citep{Na2017}.

For all present and future models the coupling between the outer corona and the lower corona is key.
This includes the coupling of the solar wind streams with the origin of CMEs.
In particular realistic boundary conditions for space weather need to consider
the origin of CMEs in the low corona and how to insert these CMEs into the solar wind.
The cone model performs this key role in the WSA-ENLIL model and has improved its predictive capabilities \citep{Dewey2015},
however it injects only a density and velocity perturbation in the solar wind,
neglecting the magnetic flux that is an essential part of the CME disturbance to space weather. 
As an alternative to this, \citet{ShiotaKataoka2016} introduced the use of a spheromak to model the perturbation from an ejecting magnetic flux rope.
It considers the injection of an idealised structure described in terms of density and magnetic field.
In \citet{Merkin2016}, where two different MHD codes are coupled, the common domain between the two codes
is in fact a time dependent boundary condition for space weather.
Other techniques to input CMEs into space weather models use coronagraph data directly,
such as in the Tappin-Howard (T-H) model \citep{TappinHoward2009a,TappinHoward2009b}.
In this model ICMEs are reconstructed from visible light images.
As an alternative, in \citet{Bisi2013} remote-sensing radio observatios of Interplanetary Scintillation (IPS)
are used to probe the inner heliosphere and to identify density irregularities in the solar wind.
When enhanced by the UCSD time-dependent tomography technique \citep{Jackson2010}.
\citet{Harrison2017} provides an extensive review on how heliospheric imaging can be used to improve
our space weather forecasts.

Many recent studies have measured our predictive capability.
\citet{ZhaoDryer2014} estimated the uncertainty in our capability to predict the arrival time of CME at 12 hours.
Also \citet{Falkenberg2011} found a similar result for WSA-ENLIL.
They explain that one of the reasons that prevents better estimates is our limited capability in reproducing
solar transients and how the CMEs are input in the background solar wind.
This seems a general trend, as no major difference has been found between the various models,
where, for instance, the DBM and ENLIL predicting capabilities differ by less than $10\%$,
except for the case of strong solar activity when ENLIL performs significantly better \citep{Vr¨nak2014}.
More recently \citet{TuckerHood2015} carried out an extensive survey that proved how much room for improvement 
there is in the field. Out of the 60 predictions made, 36 were false alarms and
when the prediction was successful the arrival time was still estimated with an error of $\sim16$ hours.
\citet{Mays2015} shows came to similar conclusions, but found that results were slightly more positive for ENLIL.
These results show that significant modelling improvements are needed to 
reduce the arrival time error and the property of the geomagnetic perturbations.
To improve these estimations more accurate boundary conditions of the 
injection of CMEs into interplanetary space are required.

In the present paper we focus on a specific kind of possible improvement:
to provide realistic and accurate boundary conditions for 
the next generation of space weather forecasting models.
In particular we need realistic initial and time dependent boundary conditions
that reflect the complexity of the transition from the solar corona to interplanetary space,
in terms of the injection of both plasma (density and velocity) and magnetic field in the solar wind.
To do this we present a novel approach where simulations of magnetic flux rope ejections
that are derived directly from surface magnetograms may be used to aid future space weather predictions.
While most magnetic structures in the 
solar corona are close to equilibrium, it is essential to identify those that lose equilibrium and then erupt.
To this end \citet{Yeates2007} have opened the way for the use of synoptic magnetograms to model the quasi-static,
non-potential evolution of the global corona. This modelling technique has proved to be accurate in predicting the 
formation and helicity of solar filaments \citep{Yeates2008}, the variation of the Sun's open magnetic flux \citep{Yeates2010JGRA} and to a lesser 
extent the location of CMEs \citep{Yeates2010}.
As already described in 
\citet{MackayVanBallegooijen2006A} and \citet{Pagano2013a}, the formation and ejection of magnetic flux ropes occur over
very different time-scales and in different dynamic regimes. The formation of magnetic flux ropes occurs slowly (days or weeks),
through a series of quasi-equilibrium states (formation times much longer than the Alfv\'en time) and in a
magnetically-dominated regime ($\beta\ll1$). In contrast, magnetic flux rope ejections takes place over a few hours, where no equilibrium
exists and where compression and heating create extended regions of high-$\beta$ plasma.
To simulate the global corona during both the formation and eruption of magnetic flux ropes,
we expand upon the work already carried out in \citet{Pagano2013a, Pagano2013b, Pagano2014}. In these studies
we have shown that coupling a quasi-static non-potential model with MHD simulations is a viable way
to describe the full life span of a flux rope: from formation to ejection. One limitation of the previous studies, was that
they only considered a small wedge-shaped portion of the Sun and simple idealised
magnetic field configurations.
We now apply the same approach to realistic observationally derived magnetic fields that cover the full sphere of the Sun as simulated in the global non-potential
model of \citet{Yeates2010}.
The model of \citet{Yeates2010} uses bipoles deduced from synoptic magnetograms to realistically model
both the time evolution of photospheric fields and the quasi-static response of coronal fields to global motions over long periods of time.
One aspect of this is the formation of magnetic flux ropes, which subsequently lose stability.
At this loss of stability we switch the modelling approach where
MPI-ARMVAC is used to follow the MHD evolution.
Throughout the MHD simulation we derive the boundary conditions for space weather forecasting tools,
i.e. the density, velocity and magnetic field distributions that the flux rope ejections inject
in the solar wind and interplanetary space.

In the present paper we present the basic physics of a model that aims to produce boundary conditions for space weather prediction models. At the present time we focus on the properties of the model, where in future studies we will consider its application and viability  in the context of an operational model along with its coupling to interplanetary models. The significance of this approach relies on its accuracy and efficiency.
The global non-potential model has been extensively tested and has been shown to be accurate in describing the evolution of the photospheric magnetic field
and subsequent formation of non-potential magnetic fields and flux ropes in the corona. General MHD simulations (capable of handling multi-$\beta$ 
domains) are the most accurate tool to model flux rope ejections and then the evolution of the flux rope through
interplanetary space. While we put forward a two-stage approach for producing the boundary condition it is of course
possible to use MHD simulations to simulate both the magnetic flux rope formation and ejections.
However, present computational power is insufficient to model the slow formation of flux ropes over days to weeks.
A MHD simulation would consume unreasonable computational resources and additionally the high number of time steps required 
to cover the physical time span would lead to 
the accumulation of significant round-off errors. Taking into account these considerations, we propose a three stage 
numerical model.
For the present paper we consider the first two steps of formation and eruption.
The third and final stage which is the connection of the ejection stage to an interplanetary evolution model 
will be carried out in the future.
This will include using the output from the MHD simulation as a time-dependent boundary condition
for driving a new generation of space weather forecasting tools. 

The structure of the paper is as follows:
in Sec.\ref{model} we present the construction of the model, in Sec.\ref{simulation} we describe in detail the MHD simulation,
in Sec.\ref{spaceweather} we explain how this study can be used in a space weather forecasting context
and we finally discuss and give conclusions in Sec.\ref{conclusion}.

\section{Model}
\label{model}
\subsection{Overview of Coupled Modelling Technique.}

In order to model the ejection of magnetic flux ropes in the global corona,
we employ a dual modelling technique of MHD simulations coupled with a quasi-static
non-potential global model \citep{Yeates2010}.
This approach is an extension of the technique successfully pioneered in 
\citet{Pagano2013a} and then further developed in \citet{Pagano2013b}
and \citet{Pagano2014}. In these studies, a magnetic configuration obtained
from the global non-potential model case study of \citet{MackayVanBallegooijen2006A}
was used  as an initial condition in the MHD simulation.
The combined technique allows us to follow the slow build-up of stress and
electric currents over observed solar timescales (days -months),
along with being able to follow the dynamic eruption timescale (min-hours).

\subsubsection{Quasi-static Model}
\label{globalmodel}
In this paper, we use a magnetic configuration obtained from a simulation run of the global
non-potential model, where the model simulated the entire time span of
Cycle 23 \citep{YeatesMackay2012}. A magnetic configuration is chosen near the end of the run, 
where the structure and
connectivity of the corona was produced by the combined effects of differential rotation, meridional 
flow, surface magnetic diffusion and magnetic flux emergence. The properties of the flux emergence events were
deduced from NSO/KP and NSO/SOLIS magnetograms and were included to maintain the accuracy
of the photospheric field compared to that found on the Sun. 
This application of the Global Model has proved to be very successful in reproducing the 
chirality of solar filaments \citep{YeatesMackay2012} throughout the solar cycle.

\begin{figure}
\centering
\includegraphics[scale=0.18,clip]{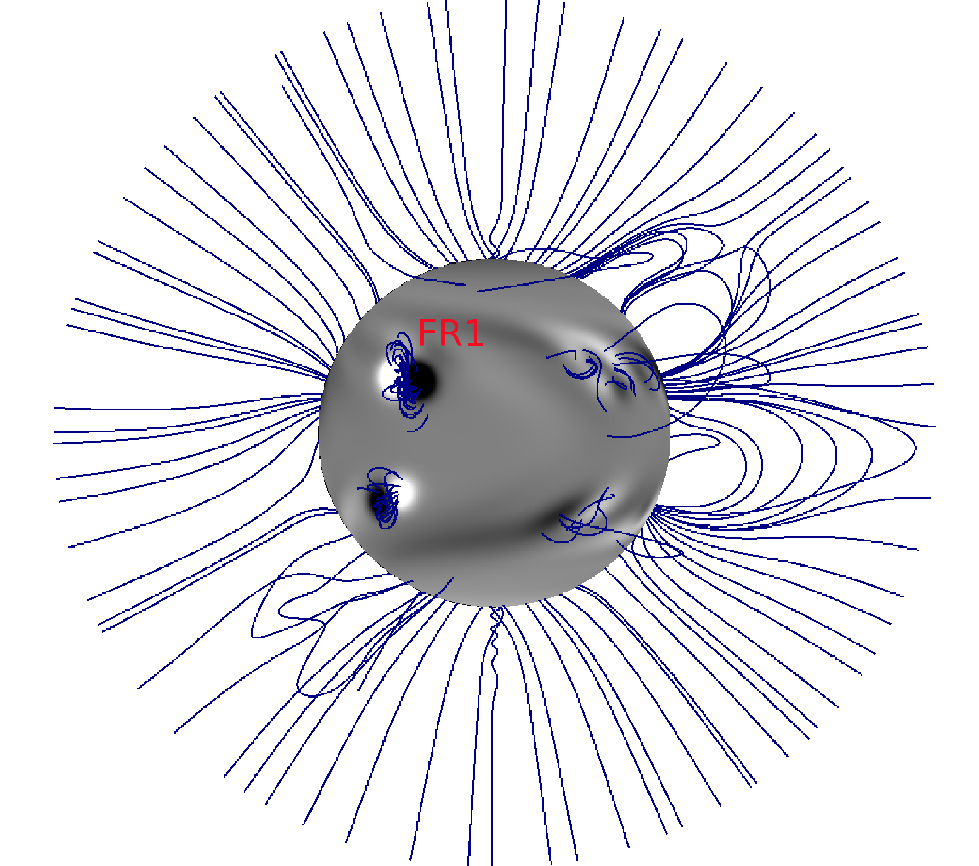}
\includegraphics[scale=0.18,clip]{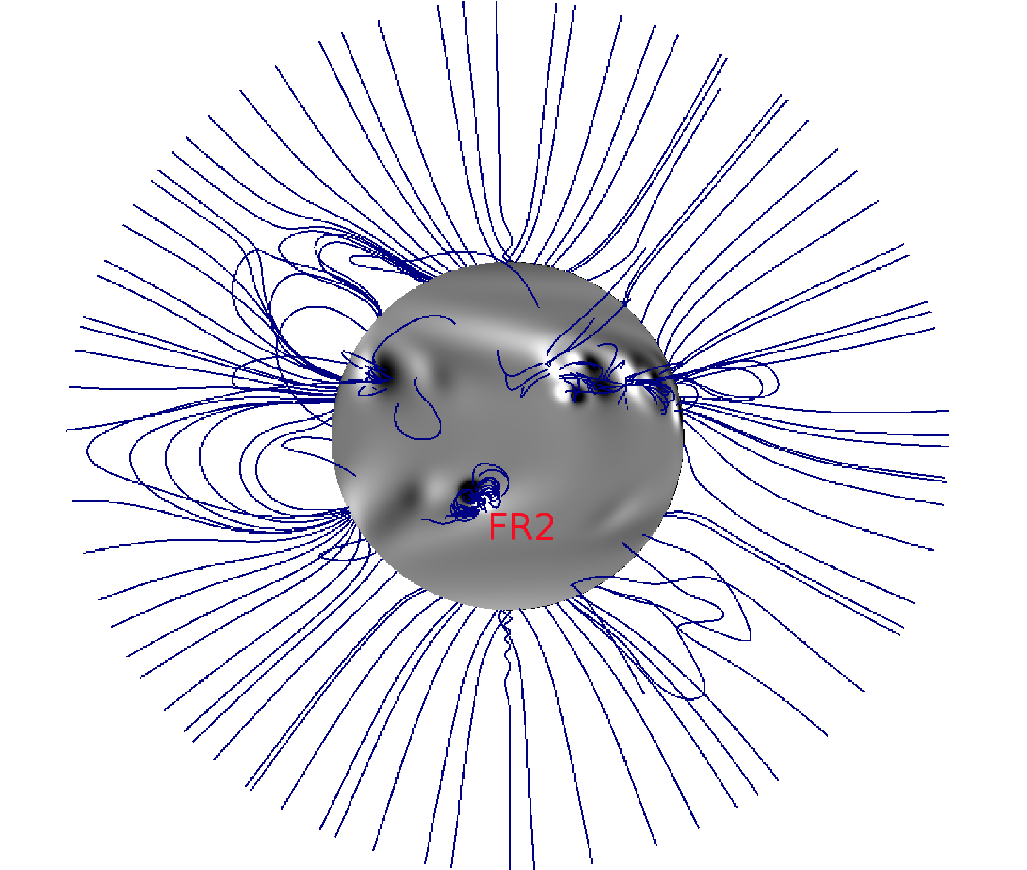}     
\caption{Representation of the magnetic configuration imported from the Global Non-potential Model.
The two panels show maps of the radial component of the magnetic field on the surface 
from two opposite points of view.
Black lines are some representative magnetic field lines to highlight existing structures.}
\label{magneticconfiguration}
\end{figure}
Additionally, we also took care by selecting
an initial condition for the MHD simulation 
where there were two relatively isolated magnetic flux ropes that had formed
and both had accumulated enough stress that they were going to erupt.
In the global model we have developed an automated technique for the identification
of erupting flux ropes. The technique is described in \citet{YeatesMackay2009}.
The global magnetic configuration under investigation can be seen from two viewpoints,
$180^{\circ}$ in longitude apart, in Fig.\ref{magneticconfiguration}.
The magnetic field lines exhibit a number of features characteristic of those
on the Sun, both at the photosphere and in the
magnetic field connectivity in the low corona.
At several locations
opposite polarity patches lie close to one-another, and these are the preferred locations of flux cancellation where flux ropes are found.
However, at only two of these locations
(FR1 in the northern hemisphere and FR2 in the southern hemisphere)
is the magnetic field sheared enough along the polarity 
inversion line to show a twisted magnetic flux rope.
At FR1 a flux rope sits above a PIL in a bipole that extends over a region of $0.4 R_{\odot} \times 0.3 R_{\odot}= 0.12 R_{\odot}^2$.
The PIL is mainly North-South directed.
At the location FR2 there is a smaller flux rope and bipole, where now the bipole extends
only for $0.23 R_{\odot} \times 0.24 R_{\odot}= 0.06 R_{\odot}^2$.
The polarities of the bipole are separated by a PIL which lies directed in a South-West to North-East direction.
For both of the active regions the flux rope lie above the PIL therefore have a slightly different orientation.
The rest of the coronal magnetic field is mostly close to potential and the other bipoles that are
present in the domain show no significant shear above their polarity inversion lines.

\subsubsection{Coupling to  MHD simulation}
To use the magnetic configuration illustrated in Fig.\ref{magneticconfiguration} as the initial condition 
for the MHD simulation, we import into the MHD simulation
all three components of the magnetic field from the global non-potential
model. A full description of how this is carried out, where both the
stability (or instability) and magnetic connectivity of the configuration is preserved, is given in
\citet{Pagano2013a} and \citet{Pagano2013b} where a number of test cases are described.

For the present paper we have slightly modified the way in which the 3D interpolation is performed. 
In spherical coordinates, let $A(r,\theta,\phi)$ be the value of a function that we want to have interpolated to the position 
$(r,\theta,\phi)$,
where $r$ is the radial distance from the centre of the Sun,
$\theta$ is the polar angle and 
$\phi$ is the azimuthal angle.
We know that it lies in the cell defined by the indexes $[i:i+1,j:j+1,k:k+1]$ 
where the original function $a$ is defined, we compute $A(r,\theta,\phi)$ as
\begin{equation}
A(r,\theta,\phi)=\sum_{i,j,k}^{i+1,j+1,k+1} a[i,j,k] V[i,j,k] / V
\end{equation}
where $V[i,j,k]$ is the volume defined by the point $(r,\theta,\phi)$ and the cell corner opposite to the 
position $[i,j,k]$ and V is the sum of all the volumes. This approach guarantees the continuity of the 
interpolated solution and its smoothness, independent of the spatial resolution of the 
grid where $A(r,\theta,\phi)$ is defined.

\subsection{MHD Simulation}

\subsubsection{Plasma distribution}
\label{plasmadistribution}

As the global non-potential model  only provides a magnetic configuration, to produce a complete set of 
MHD variables in the initial condition we need to determine distributions of plasma density, velocity and temperature. In 
specifying these we 
aim at a realistic and general representation of the solar corona, where the distribution of plasma takes into 
account the heterogeneity of the structures of the solar corona. This heterogeneity can be simplified by
representing the solar corona as a combination of
dense active regions and less dense quiet Sun regions.
In addition, there are dominantly horizontal magnetic fields (such as prominences or
filaments) which are generally two orders of magnitude cooler and denser than typical coronal values.

We define the following proxy function to link the plasma temperature and density
to the magnetic field:
\begin{equation}
\label{omegab}
\omega=\sqrt{\omega_r^2+\omega_{\theta}^2+\omega_{\phi}^2}
\end{equation}
where
\begin{equation}
\label{omegarthetaphi}
\omega_r=\frac{\left|\vec{B}\times\vec{\nabla B_r}\right|^2}{\left|\vec{\nabla B_r}\right|^2}, ~~
\omega_{\theta}=\frac{\left|\vec{B}\times\vec{\nabla B_{\theta}}\right|^2}{\left|\vec{\nabla B_{\theta}}\right|^2}, ~~
\omega_{\phi}=\frac{\left|\vec{B}\times\vec{\nabla B_{\phi}}\right|^2}{\left|\vec{\nabla B_{\phi}}\right|^2}
\end{equation}

The function $\omega$ is positive definite and peaks where the magnetic field exhibits a complex twisted
field, e.g. near the axis of a magnetic flux rope. As  the value of $\omega$ is also proportional to the
magnetic field intensity, it is higher near the solar surface (where the magnetic field is more intense)
and lower at further radial distances from the solar surface.

In order to effectively use $\omega$ to model the solar atmosphere we define the functions:
\begin{equation}
\Omega=\frac{\Omega_B+\Omega_{\theta}}{2}+\frac{\left|\Omega_B-\Omega_{\theta}\right|}{2},
\end{equation}
\begin{equation}
\Omega_\theta(\theta)=\frac{\arctan\left(\frac{\theta-(\pi-\theta^{\star})}{\Delta\theta}\right)}{\pi}-\frac{\arctan\left(\frac{\theta-\theta^{\star}}{\Delta\theta}\right)}{\pi}+1,
\end{equation}
\begin{equation}
\Omega_B(\omega)=\frac{\arctan{\left(\frac{\omega-\omega^{\star}}{\Delta\omega}\right)}}{\pi}+0.5,
\end{equation}
where $\Omega_{\theta}$ and $\Omega_B$ are functions bound between $0$ and $1$
while $\Omega$ is defined to pick the higher between $\Omega_{\theta}$ and $\Omega_B$.
$\Omega_{\theta}$ is a function that only depends on the coordinate $\theta$ that defines two regions
near the poles where we quench any dynamics close to the boundaries of the simulation.
Using $\Omega$ the temperature is defined by:
\begin{equation}
\label{tempomega}
T=\Omega(T_{flux rope}-T_{corona})+T_{corona}.
\end{equation}

Next, the thermal pressure is independently specified by the solution for hydro-static equilibrium
with a uniform temperature set equal to $T_{corona}$,
\begin{equation}
\label{pressurestratification}
p=\frac{\rho_{LB}}{\mu m_p}k_B 2 T_{corona} \exp\left({-\frac{M_{\odot}G \mu m_p}{2 T_{corona} k_B R_{\odot}}}\right) \exp\left({\frac{M_{\odot}G \mu m_p}{2 T_{corona} k_B r}}\right),
\end{equation}
where $\rho_{LB}$ is the density at $r=R_{\odot}$ when $|B|=0$,
$\mu=1.31$ is the average particle mass in the solar corona, $m_p$ is the proton mass and
$k_B$ is Boltzmann constant.
Finally, the density is simply given by the equation of state applied to Eq.\ref{tempomega} and
 Eq.\ref{pressurestratification}:
\begin{equation}
\label{eos}
\rho=\frac{p}{T(\vec{B})}\frac{\mu m_p}{k_B}.
\end{equation}
This produces an inhomogeneous solar corona of cool dense flux ropes and hotter emptier corona arcades.

\subsubsection{MHD simulation}
Using the approach described in Sec.\ref{plasmadistribution}, we construct the initial condition for the
MHD simulation, where Table~\ref{tableparameters} shows the value used in our model for all parameters.
\begin{table}
\caption{Parameters}             
\label{tableparameters}      
\centering                          
\begin{tabular}{c c c}        
\hline\hline                 
Parameter & value & Units  \\    
\hline                        
   $\rho_{LB}$ & $2.77\times10^{-15}$ & $g/cm^3$  \\      
   $\theta^{\star}$ & $0.13$  & radiants \\
   $\Delta\theta$ & $0.005$ & radiants \\
   $\omega^{\star}$ & $30$  & $G$ \\
   $\Delta\omega$ & $3$  & $G$ \\
   $T_{flux rope}$ & $10^4$ & K \\
   $T_{corona}$ & $2\times10^6$ & K \\

\hline                                   
\end{tabular}
\end{table}
With this initial condition we use the MPI-AMRVAC software~\citep{Porth2014}, to solve the MHD equations
where external gravity is included as a source term,
\begin{equation}
\label{mass}
\frac{\partial\rho}{\partial t}+\vec{\nabla}\cdot(\rho\vec{v})=0,
\end{equation}
\begin{equation}
\label{momentum}
\frac{\partial\rho\vec{v}}{\partial t}+\vec{\nabla}\cdot(\rho\vec{v}\vec{v})
   +\nabla p-\frac{(\vec{\nabla}\times\vec{B})\times\vec{B}}{4\pi}=\rho\vec{g},
\end{equation}
\begin{equation}
\label{induction}
\frac{\partial\vec{B}}{\partial t}-\vec{\nabla}\times(\vec{v}\times\vec{B})=0,
\end{equation}
\begin{equation}
\label{energy}
\frac{\partial e}{\partial t}+\vec{\nabla}\cdot[(e+p)\vec{v}]=\rho\vec{g}\cdot\vec{v},
\end{equation}
where $t$ is time, $\rho$ is density, $\vec{v}$ velocity, $p$ thermal pressure and $\vec{B}$ the magnetic field.
The total energy density $e$ is given by
\begin{equation}
\label{enercouple}
e=\frac{p}{\gamma-1}+\frac{1}{2}\rho\vec{v}^2+\frac{\vec{B}^2}{8\pi}
,\end{equation}
where $\gamma=5/3$ denotes the ratio of specific heats. The expression for solar gravitational acceleration is
given by
\begin{equation}
\label{solargravity}
\vec{g}=-\frac{G M_{\odot}}{r^2}\vec{\hat{r}},
\end{equation}
where $G$ is the gravitational constant, $M_{\odot}$ denotes the mass of the Sun,
and $\vec{\hat{r}}$ is the unit vector.

The computational domain is composed of $256\times256\times512$ cells, distributed on a uniform spherical grid.
With this resolution the simulation domain extends over $3$ $R_{\odot}$ in the radial direction starting from
$r=R_{\odot}$. The co-latitude, $\theta$, spans from $\theta=0.75^{\circ}$ to $\theta=179.25^{\circ}$ and the 
longitude, $\phi$, spans $360^{\circ}$. The boundary conditions are treated with a system of ghost cells
and match those used in \citet{Yeates2010}. Open boundary conditions are imposed at the outer boundary, 
reflective boundary conditions are set at the $\theta$ boundaries and the $\phi$ boundaries are periodic.
The reflective $\theta$ boundary condition does not allow any plasma or magnetic flux to pass through.
The stability at this boundary is reinforced through the use of the function $\Omega_{\theta}$, which
results in plasma near the $\theta$ boundary having a density significantly higher than that prescribed by 
gravitational stratification. This enhancement quenches any upward motion at this boundary. At the lower boundary
which represents the photosphere, we impose a fixed boundary condition taken from the first two $\theta$-$\phi$
planes of cells derived from the global non-potential model.

The interpolation technique used to import the magnetic field configuration from the spatial grid of the 
global non-potential model, to the alternative grid that we use in the MHD simulation only applies to 
$r<2.5 R_{\odot}$. In the region beyond $r=2.5 R_{\odot}$ we assume a purely radial field where magnetic flux 
is conserved:
\begin{equation}
\label{brover25r}
B_r(r>2.5 R_{\odot},\theta,\phi)=B_r(2.5 R_{\odot},\theta,\phi)\frac{\left(2.5 R_{\odot}\right)^2}{r^2}.
\end{equation}

\begin{figure}
\centering
\includegraphics[scale=0.28,clip]{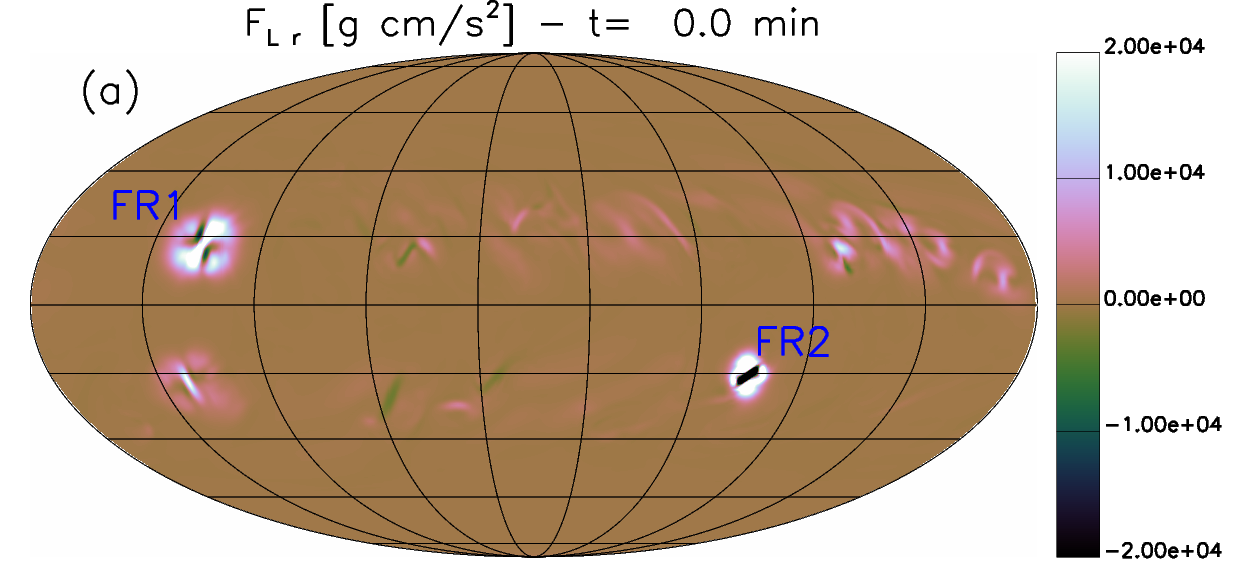}
\includegraphics[scale=0.28,clip]{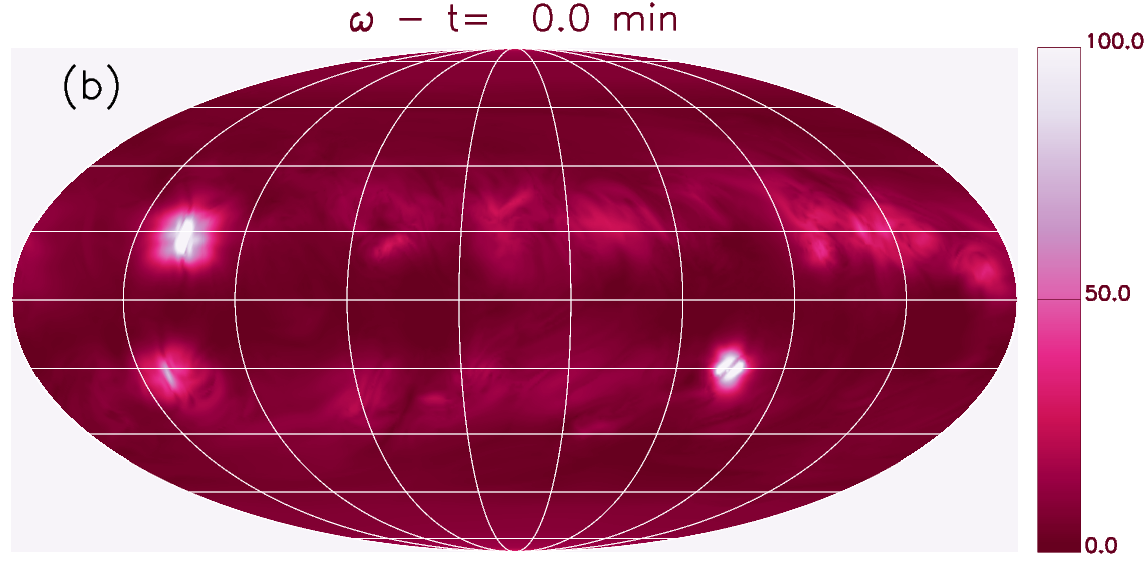}
\includegraphics[scale=0.28,clip]{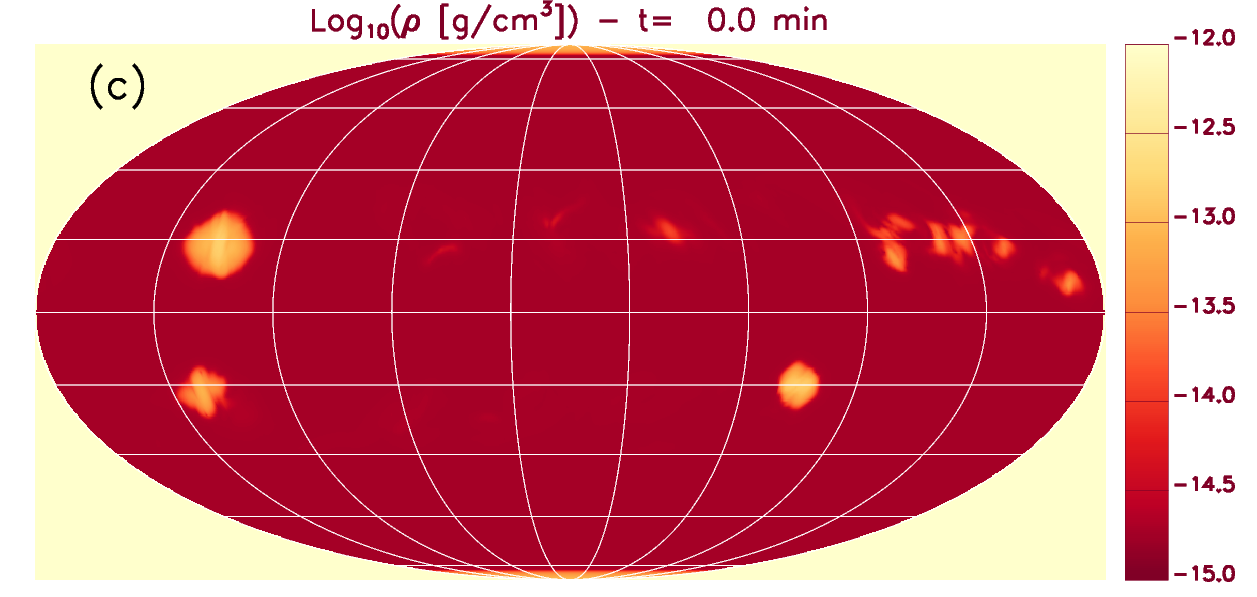}
\includegraphics[scale=0.28,clip]{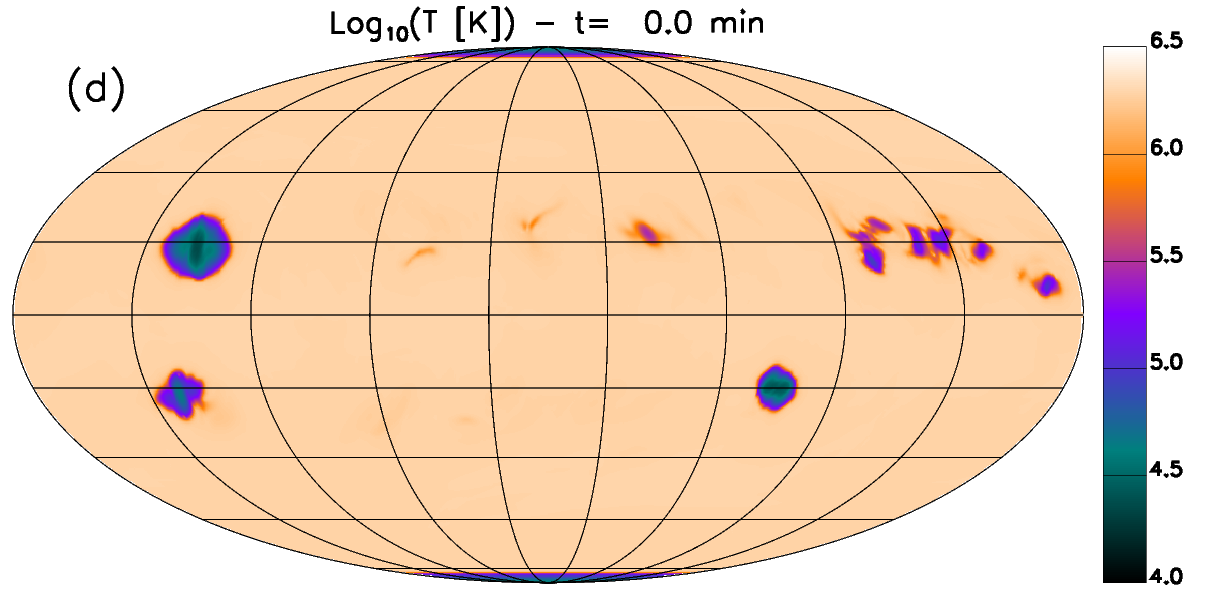}
\caption{Mollweide projections (central meridian at longitude $\phi=180^{\circ}$) on the lower boundary of the MHD simulation of
 (a) Lorentz Force,  (b) $\omega$, (c) $Log_{10}(\rho [g/cm^3])$ and (d) $Log_{10}(T [K])$.}
\label{initialcond}
\end{figure}

The key properties of the initial condition in the MHD simulation can be seen in Fig.\ref{initialcond} which
shows in Mollweide projection distributions of (a) the radial component of the Lorentz force at the lower
 boundary, 
(b) the function $\omega$ derived from the magnetic configuration and finally the resulting distributions of 
(c) plasma density and (d) temperature.
The Mollweide projection has the advantage of showing in a single field of view
the whole solar disk and therefore clearly illustrates the connectivity between different regions.
The Lorentz force exhibits several areas of strong outward radial force 
shown as the white regions. In particular 
two structures show a significantly larger positive Lorentz force compared to surrounding areas. 
One lies in the northern hemisphere to the left of central meridian and the other is in the southern
hemisphere to the right of central meridian.
These are
at locations where large flux ropes have formed and can no longer be fully contained by the overlying fields.
Close to where the flux ropes lie, with their positive radial component of Lorentz force, smaller regions 
where the Lorentz force is negative can be seen.

The map of $\omega$ (Fig.\ref{initialcond}b) follows a similar distribution, where several small structures 
are visible along the active latitudes. Three zones exist where $\omega$ is significantly larger 
with peak values nearly $\sim100$ times greater than ambient background values.  
Flux ropes show higher values of $\omega$, as close to flux ropes all the three terms 
$\omega_r$, $\omega_{\theta}$, and 
$\omega_{\phi}$ are important at different locations. Namely $\omega_r$ peaks at 
the center of the magnetic flux rope axis, while $\omega_{\theta}$ and $\omega_{\phi}$ peak near the footpoints.
Finally, the maps of density (Fig.\ref{initialcond}c) and temperature (Fig.\ref{initialcond}d) show that the 
locations with a more complex magnetic configuration are denser and cooler. With this the 
magnetic flux ropes in the simulation appear as regions with
density and temperature one or two orders of 
magnitude denser and colder than the surrounding ambient values. Thus the initial conditions produce an 
inhomogeneous corona that exhibits many features found on the Sun.

The initial plasma $\beta$ in the simulation ranges between $\beta\sim10^{-3}$ 
at the flux ropes to  $\beta\sim1$ in confined regions where the magnetic 
field is weak. While there is a wide range, normally the value of $\beta$ lies between $0.1$ and $0.01$ throughout the 
vast majority of the volume. Due to the low $\beta$ values at the flux ropes, the strongest unbalanced force 
in the initial 
condition is the radially directed Lorentz force at these locations. 
In addition to the unbalanced Lorentz force the radial profile of 
density and pressure also does not prescribe a balance between the thermal pressure gradient and gravity.
While this pressure force imbalance exists, the consequences have been discussed in detail in \citet{Pagano2013b}, 
where it is shown that any evolution due to these imbalances occurs over timescales much longer 
than that of the eruption dynamics triggered by the Lorentz force. As such we can neglect these 
effects.

\section{Simulation}
\label{simulation}

\begin{figure}
\centering
\includegraphics[scale=0.27,clip]{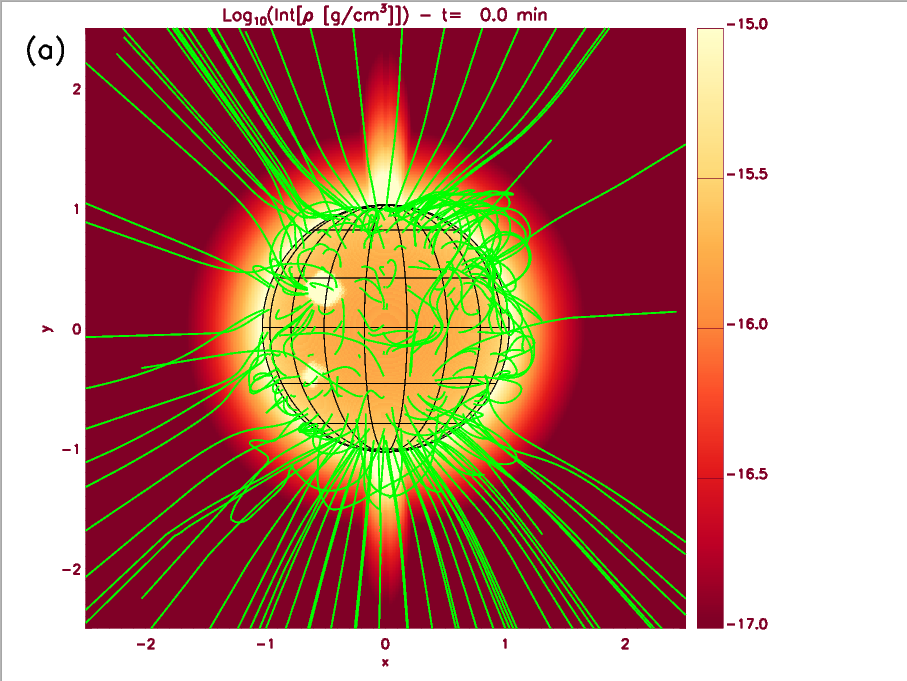}
\includegraphics[scale=0.27,clip]{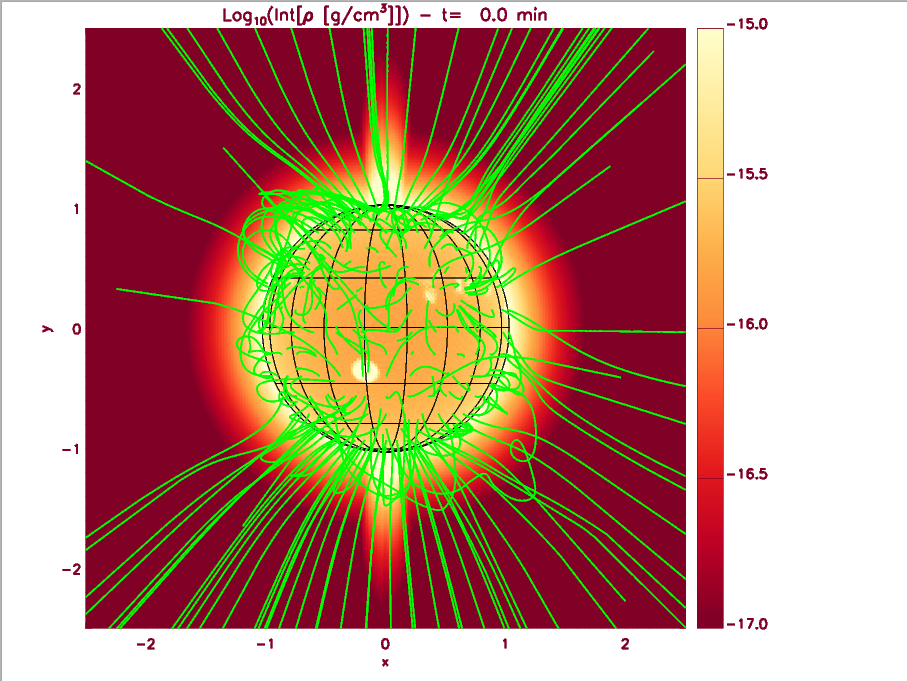}

\includegraphics[scale=0.27,clip]{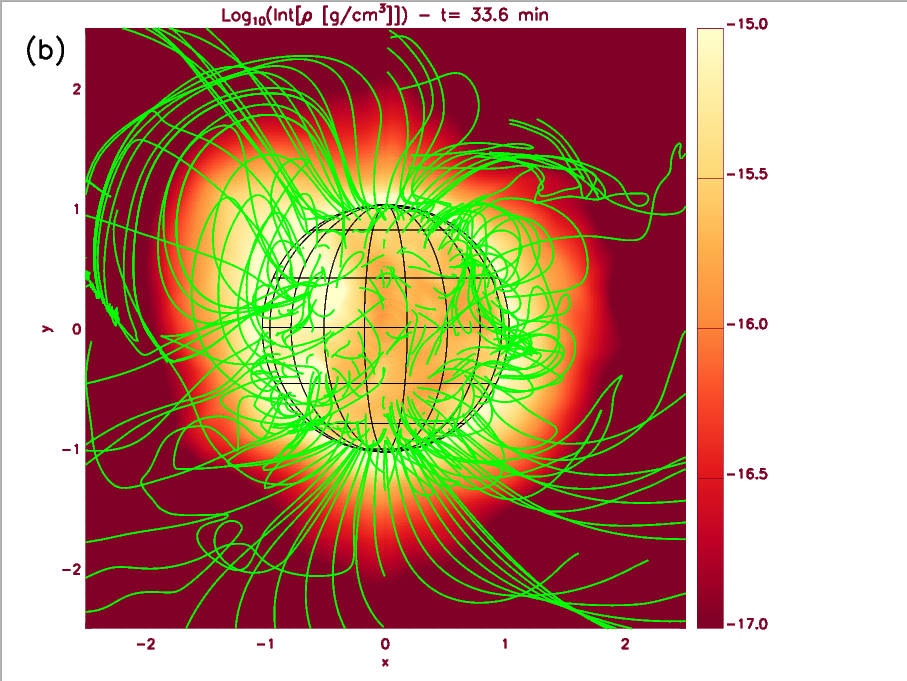}
\includegraphics[scale=0.27,clip]{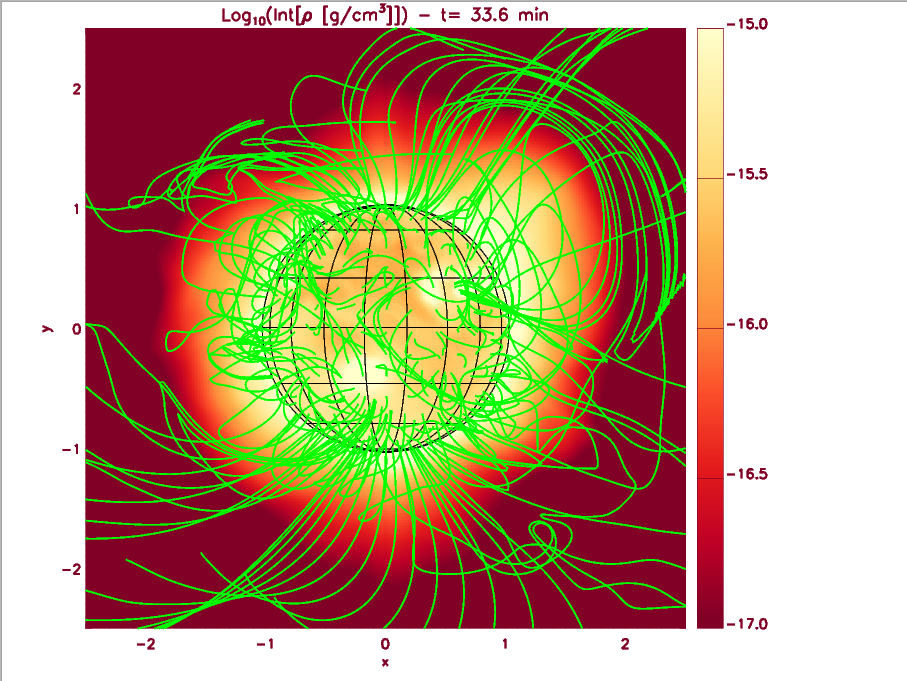}

\includegraphics[scale=0.27,clip]{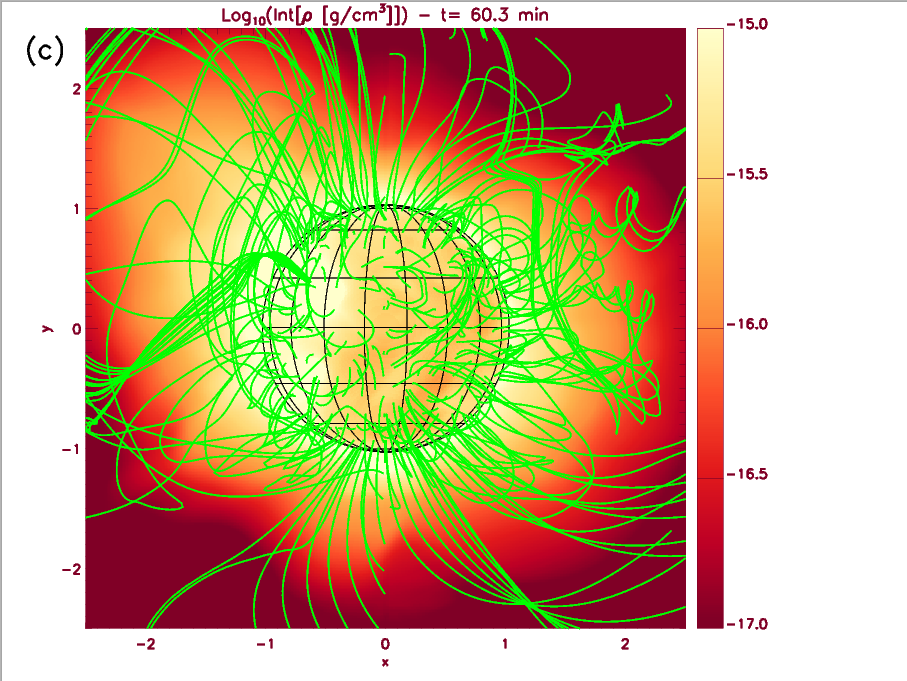}
\includegraphics[scale=0.27,clip]{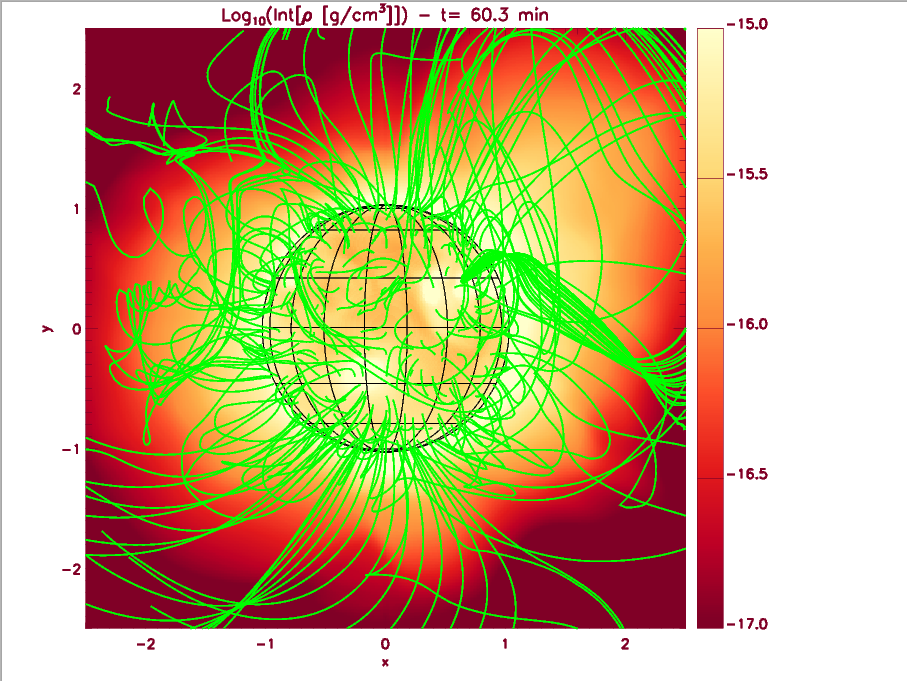}
\caption{Maps of the column density seen from either side of the Sun in the MHD simulation 
along with some superimposed magnetic field lines traced from the lower boundary (green) 
at  $t=0$ $min$ (a), $t=33.6$ $min$ (b), and $t=60.3$ $min$ (c).
Central meridian for left-hand side panels is at $\phi=90^{\circ}$ 
and at $\phi=270^{\circ}$ for right-hand sided panels
and the FOVs are chosen to be same as in Fig.\ref{magneticconfiguration}}
\label{fulldisk}
\end{figure}

The MHD simulation produces two individual eruptions that originate from the two twisted magnetic flux ropes 
shown in Fig.\ref{magneticconfiguration} and Fig.\ref{initialcond}.
In each case, the radial Lorentz force excess due to a local non-equilibrium, triggers 
the dynamic evolution of the eruption that occurs in the simulation.
Fig.\ref{fulldisk} shows images of the column density where the central meridian longitude is chosen to be
$\phi=90^{\circ}$ (left-hand side column) and $\phi=270^{\circ}$ (right-hand side column)
at (a) $t=0$ $min$, (b) $t=33.6$ $min$ and (c) 
$t=60.3$ $min$ where magnetic field lines (green) have been drawn from the photospheric boundary.
The two flux ropes are located such that one is in the northern hemisphere, while the other is in the 
southern hemisphere, and they lie approximately $180^{\circ}$ in longitude ($\phi$) apart.
When the ejections occur they 
displace dense plasma to larger radii, which results in an increased column density around the locations of 
each 
flux rope. This increased density can be seen as the two circular structures expanding at either limb, 
one towards North and $\phi=0^{\circ}$, the other towards South and $\phi=180^{\circ}$.
During the ejections the global magnetic field undergoes a rapid evolution.
Even though there is a rapid evolution, it is possible to follow the evolution of the twisted magnetic 
field lines of the flux ropes. After  $t=60.3$ $min$ the two circular expanding structures of density
have expanded sufficiently that they cross the outer boundary of the domain.

\begin{figure}
\centering
\includegraphics[scale=0.40,clip]{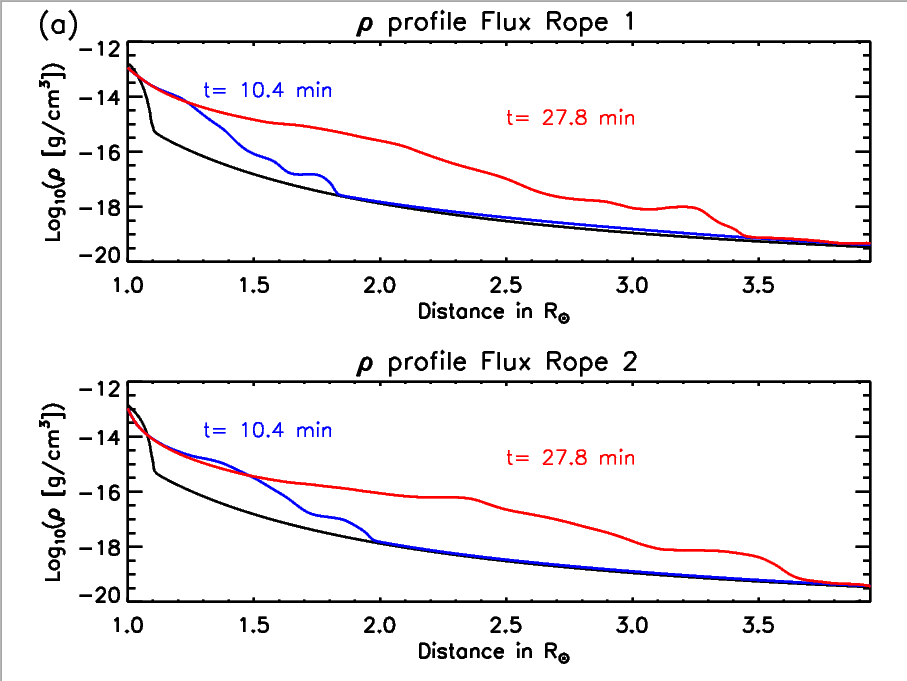}

\includegraphics[scale=0.40,clip]{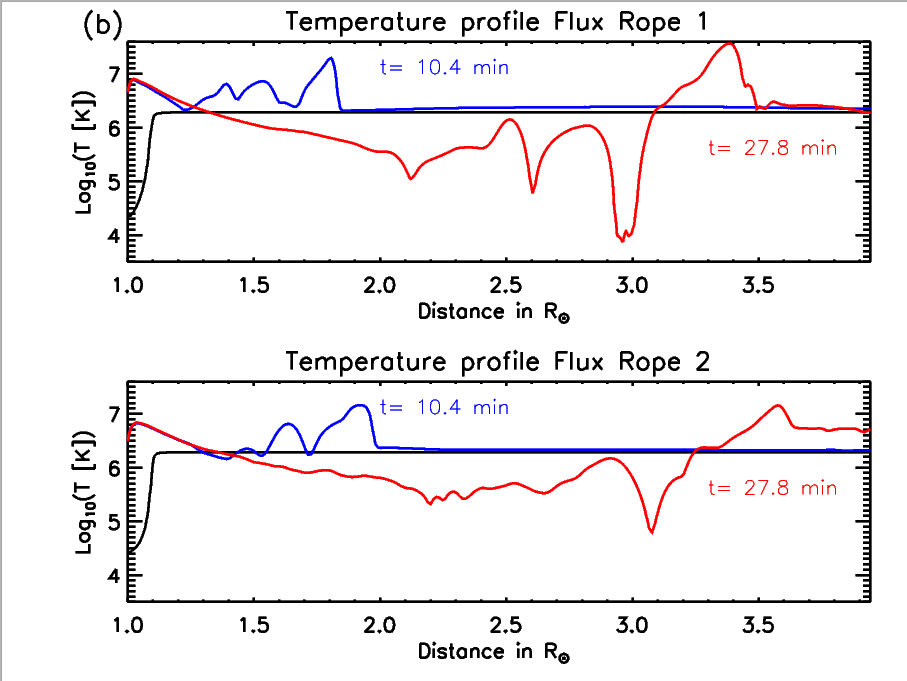}
\caption{(a) Cuts of $Log_{10}(\rho [g/cm^3])$ along the radial direction from solar surface to the outer boundary 
at the locations of FR1 and FR2 at $t=10.4$ $min$ (blue line) and $t=27.8$ $min$ (red line).
(b) Cuts of $Log_{10}(T [K])$ along the radial direction from solar surface to the outer boundary 
at the locations of FR1 and FR2 at $t=10.4$ $min$ (blue line) and $t=27.8$ $min$ (red line).}
\label{radialcutsrhotemp}
\end{figure}

\begin{figure}
\centering
\includegraphics[scale=0.40,clip]{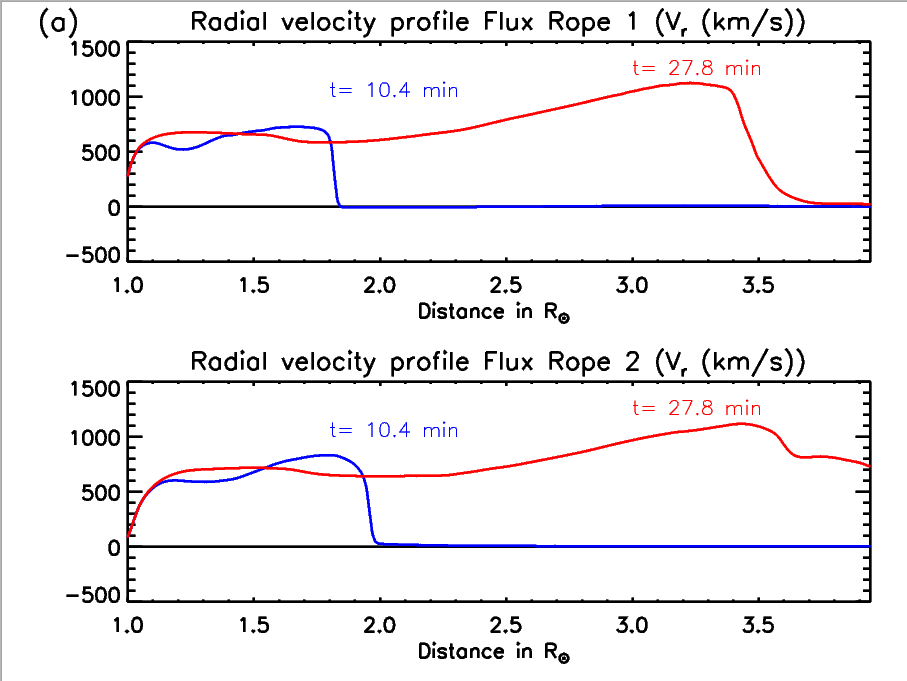}

\includegraphics[scale=0.40,clip]{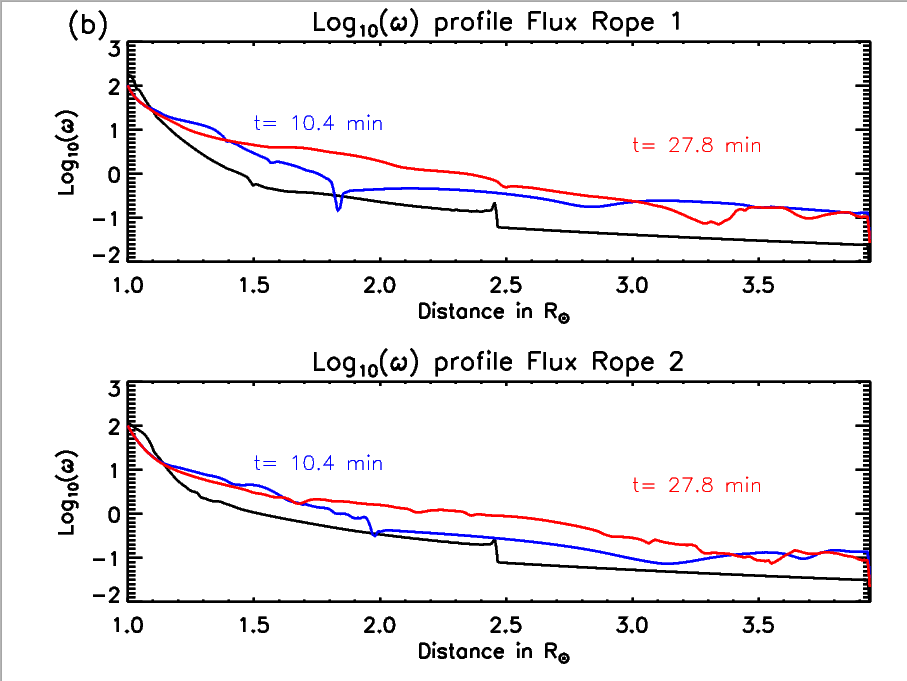}
 
\caption{(a) Cuts of radial velocity along the radial direction from solar surface to the outer boundary 
at the locations of FR1 and FR2 at $t=10.4$ $min$ (blue line) and $t=27.8$ $min$ (red line).
(b) Cuts of $Log_{10}(\omega)$ along the radial direction from solar surface to the outer boundary 
at the locations of FR1 and FR2 at $t=10.4$ $min$ (blue line) and $t=27.8$ $min$ (red line).}
\label{radialcutsvz}
\end{figure}

To better understand the evolution of the two flux rope ejections we plot radial cuts of
density, temperature, radial velocity and $\omega$ from the photospheric boundary to the outer boundary 
(Fig.\ref{radialcutsrhotemp} and Fig.\ref{radialcutsvz}). These plots are taken
along a radial line
intersecting with the initial position of the centers of the two magnetic flux ropes.
Fig.\ref{radialcutsrhotemp}(a) shows the density at $t=0$ $min$ (black line), $t=10.4$ $min$ 
(blue line) and $t=27.8min$ (red line) for both Flux Rope 1 and 2. In each plot the black line shows the initial density profile with the
dense flux rope at low heights along with the stratified atmosphere. At later times the presence 
of the ejection is identified by an excess in density compared to that of the initial stratified atmosphere. 
At $t=10.4$ $min$ the density excess of the ejected plasma has reached, $\sim 1.8$ $R_{\odot}$ for FR1 
and $\sim 2$ $R_{\odot}$ for FR2 showing that they are traveling at different speeds.
Similar estimations can be made with the fronts
of the two ejections and at different times.
In all cases, we consistently find that FR2 travels about $25\%$ faster than FR1.

At the lead front of the density excess 
there is a very sharp temperature increase to $T\sim 10^7$ $K$ (Fig.\ref{radialcutsrhotemp}b). 
Beneath the front the temperature profile also shows significant structure.
At $t=10.4$ $min$ the temperature profile in the atmosphere exhibits a dip at the location of the axis for both
FR1 ($1.2 R_\odot$) and FR2 ($1.4 R_\odot$). It however should be noted that the temperature at the center of the flux rope is comparable
to that of the background external corona at $2 MK$. While this value is high, it is still
cooler than that of the surrounding ejection.
At $t=10.4$ $min$ the radial velocity profiles in Fig.\ref{radialcutsvz}(a) for both FR1 and FR2 exhibit a very 
similar  behavior, where there is an initial sharp rise at the photosphere, followed by a slight dip before becoming
roughly level between 1.5 - 2$R_\odot$. 
In both cases at the lead edge there is a sharp fall to zero. Combining the information in 
Fig.\ref{radialcutsrhotemp}
and Fig.\ref{radialcutsvz} it can be seen that the ejection of both flux ropes can be characterized by a front 
(density excess, sharp increase in temperature and radial velocity) and a core (density excess, temperature dip, 
slower than the front). While this is characteristic of the early stages of the ejection similar structures are visible 
in the plots at $t=27.8$ min. However, there are some differences from the earlier times which include: i) the volume of
density excess increases and thus the density at the flux rope decreases, ii) the temperature at the front increases, 
while the temperature 
dips increase in number and become colder and iii) the radial velocity at the front slightly increases and with 
this the speed 
of the flux rope ejection. In addition to the features above, we also find that for both FR1 and FR2 $\omega$ shows an 
increase in value with respect to that of the initial condition.
This occurs both in the region beyond the front but  more prominently near the location of the flux rope center.

\begin{figure}
\centering
\includegraphics[scale=0.50,clip]{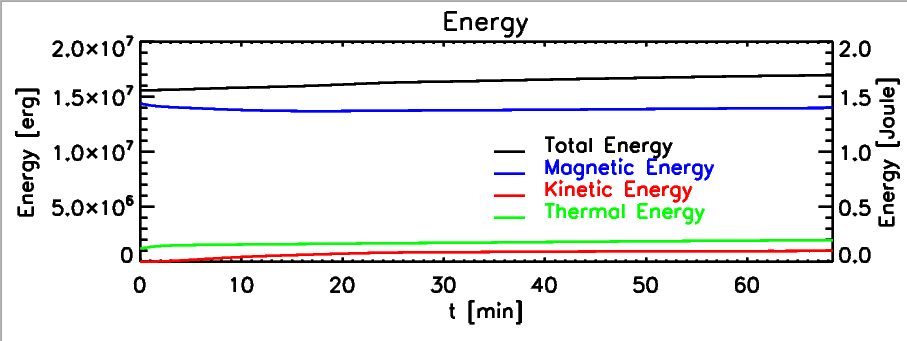}
\caption{Energy in the MHD simulation integrated over the whole spatial domain as a function of time.
Black line shows total energy, blue magnetic energy, red kinetic energy, and green thermal energy.}
\label{energybox}
\end{figure}

The  occurrence of the two ejections has a significant effect on the energy budget of the solar corona.
Fig.\ref{energybox} shows the total energy as a function of time in the MHD simulation (black line). The total energy 
is approximately conserved and after $t=68.44$ $min$ the total energy has increased by less than $9\%$.
The increase can be accounted for as there is
a net flux of energy into the simulation after taking into account the energy flux across the lower and outer 
boundaries. At the start, the magnetic energy accounts for about $90\%$ of the total energy and the thermal energy 
for the  remaining $10\%$. By the end of the simulation the kinetic energy is $6\%$ of the total energy where its 
increase is due to
the conversion of magnetic energy which falls to $80\%$ of the total energy, while the thermal energy slightly 
increases.

\section{Toward a space weather application}
\label{spaceweather}

One of the goals of the present work is to provide accurate boundary conditions of the 
outer solar corona that can be used in future space weather forecasting tools, such as those
for the solar wind and the evolution of Interplanetary Coronal Mass Ejections (ICMEs). 
To achieve a realistic model of space weather conditions at 1 AU,
it is key that accurate initial conditions for the injection of plasma and 
magnetic flux into the solar wind are specified close to the Sun. These boundary conditions in turn require accurate
time-dependent modelling of low coronal magnetic fields during the build up to
and occurrence of an eruption.

Our technique of coupling the global non-potential model with the MHD numerical solver MPI-AMRVAC
is suitable for space weather forecasting purposes because it accurately models observed magnetic field
configurations on the Sun and is very computationally efficient. It allows us to model the fast dynamic 
ejection of magnetic flux ropes, maintaining the accuracy and generality of a full MHD model, while
allowing us to model the slow quasi-static formation of flux ropes and the global corona in a numerically 
cheaper way. Initial estimations suggest that the global non-potential model is between $10^4$ and 
$10^5$ times faster than the MPI-AMRVAC in advancing the simulation in physical time. This therefore allows for the
near real time simulation of observed magnetic fields on the Sun.
We now discuss how the dual approach described above may provide useful boundary
conditions for space weather models.

\subsection{Injection of the CME into interplanetary space}
In this section, we present some exploratory results on how  our combined model injects 
a CME into interplanetary space. As we have set the outer boundary 
condition at $4$ $R_{\odot}$ we focus on the conditions of the plasma and magnetic 
field as a consequence of the eruption before it starts to interact with the solar wind.

\begin{figure}
\centering
\includegraphics[scale=0.30,clip]{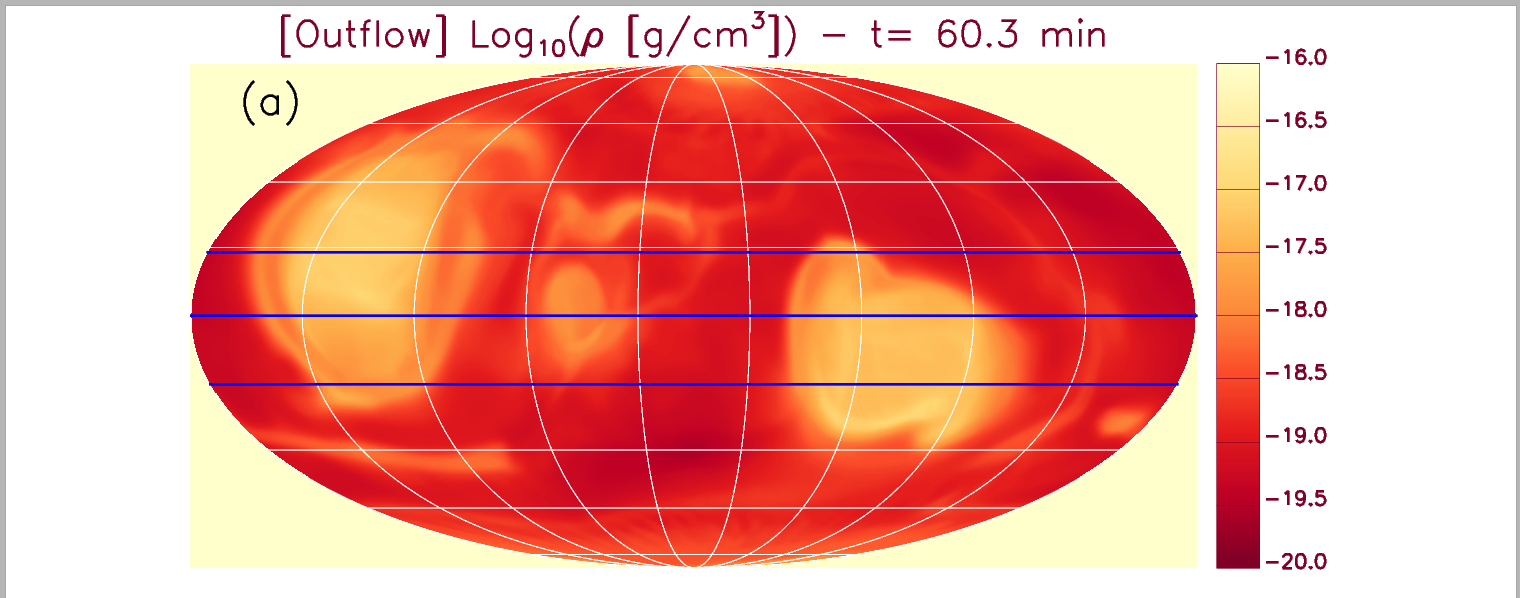}

\includegraphics[scale=0.30,clip]{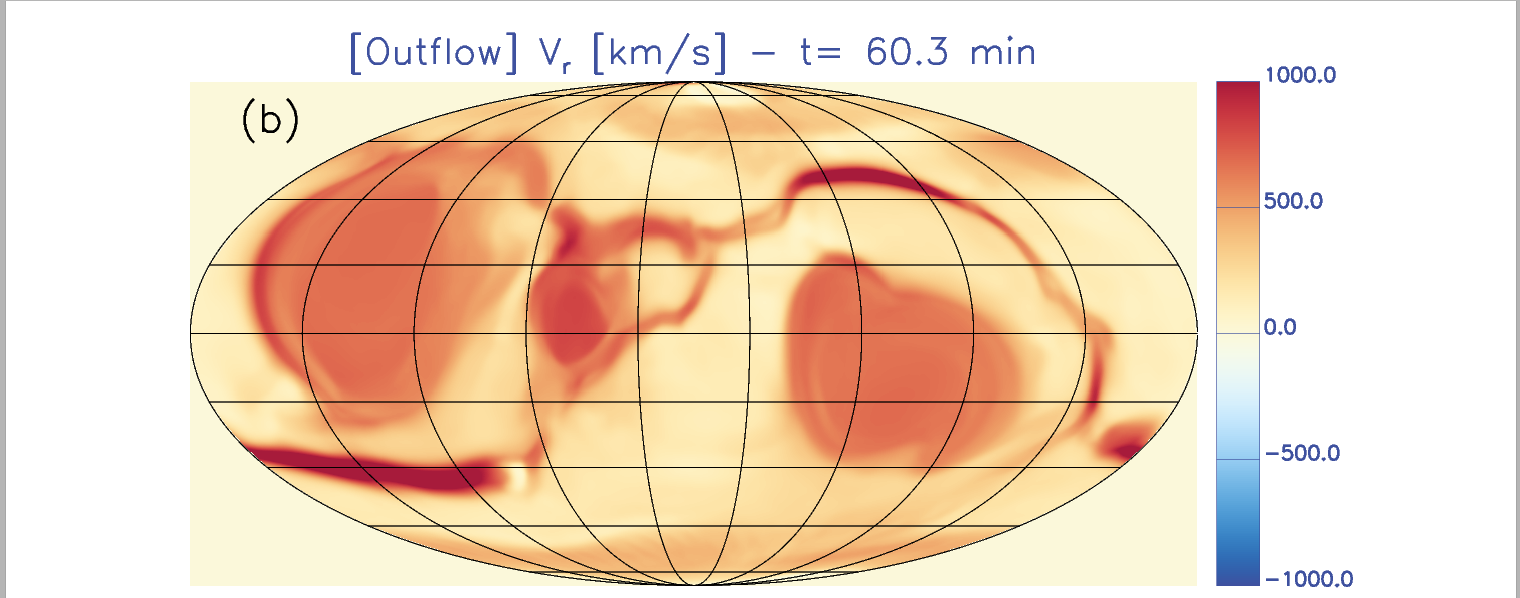}
\caption{Mollweide projections (central meridian at longitude $\phi=180^{\circ}$) at the outer boundary of the MHD simulation at $t=60.3$ $min$
of (a) $\rho [g/cm^3]$ and (b) radial velocity of the plasma $[cm/s]$.}
\label{outplasmamaps}
\end{figure}

Fig.\ref{outplasmamaps} shows Mollweide projections of (a) density and (b) the radial velocity at
the outer boundary of the simulation at a time near when both flux ropes cross this boundary ($t=60.3$ $min$).
The two flux rope ejections appear as density excesses covering an area of some tens of squared degrees
where FR1 involves a wider area than FR2.
One predominately lies in the northern hemisphere and the other in the southern hemisphere, but both
straddle the equator. The density of the flux ropes is around $100$ times larger than that of the background 
corona, which has a density of around $\sim10^{-19}$ $g/cm^3$. Due to the two orders of magnitude difference
it appears that at the outer boundary the flux rope has a near homogeneous density. However, on closer 
inspection of the data internal variations in the density
are also visible,
where higher density structures lie in the centre of the ejection for FR1,
and at the southern boundary for the ejection of FR2.
A similar pattern can also be seen in the radial velocity 
maps, where both ejections exhibit an outward velocity in the order of hundreds of $km/s$.
Outside of the region corresponding to the erupting flux ropes the corona has a significantly
smaller outward velocity.
It should  be noted that in this simulation the fastest speeds are not found at the flux rope locations,
but rather along the inversion line at the outer boundary,
where it is possible that the eruptions trigger some magnetic reconnection processes that accelerate the plasma.
At this location where the radial component 
of the magnetic field changes sign, the radial outward velocity reaches $\sim1000$ $km/s$.
This outflow is 
related to a streamer-like phenomenon, where plasma
accelerates in conjunction with reconnection at a current sheet.
This is partly due to the magnetic field from the global simulation 
relaxing at this location in the high corona when the MHD simulation starts.
This is an artefact of the coupling of the two codes 
that we will tackle in the future.
Whilst this occurs, the low density associated with this feature means that it does not contribute significantly to the momentum flux
and is essentially negligible.
The internal structure of the radial velocity patterns at both flux ropes differs.
For FR1 there are higher radial velocities at the borders of the region involved in the ejection
and the region of high positive radial velocity extends to near the polarity inversion line .
In contrast the radial velocity pattern of FR2 shows more modest internal structuring and 
remains isolated from the polarity inversion line.

\begin{figure}
\centering
\includegraphics[scale=0.38,clip]{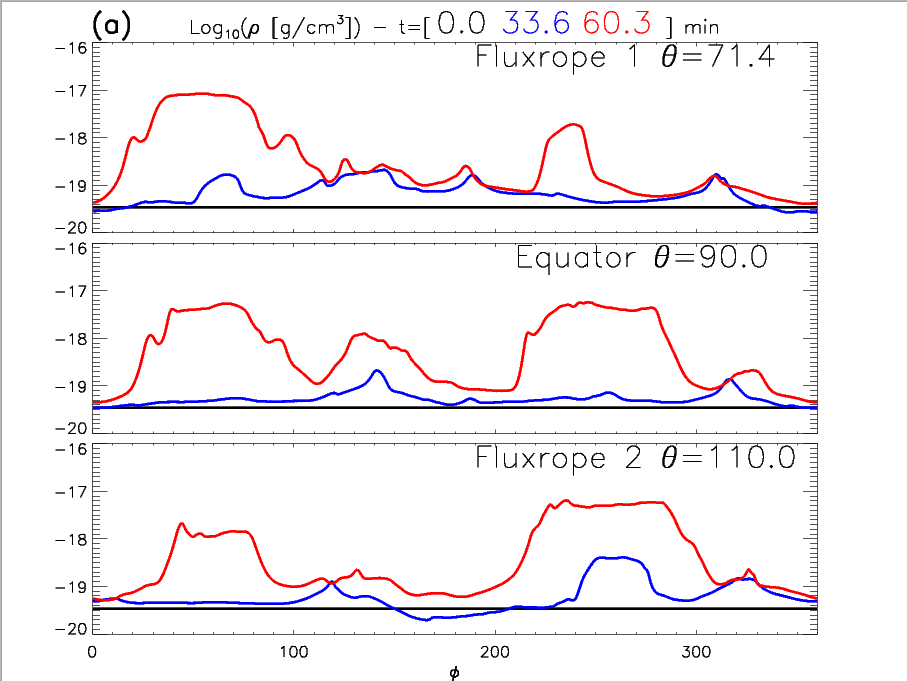}

\includegraphics[scale=0.38,clip]{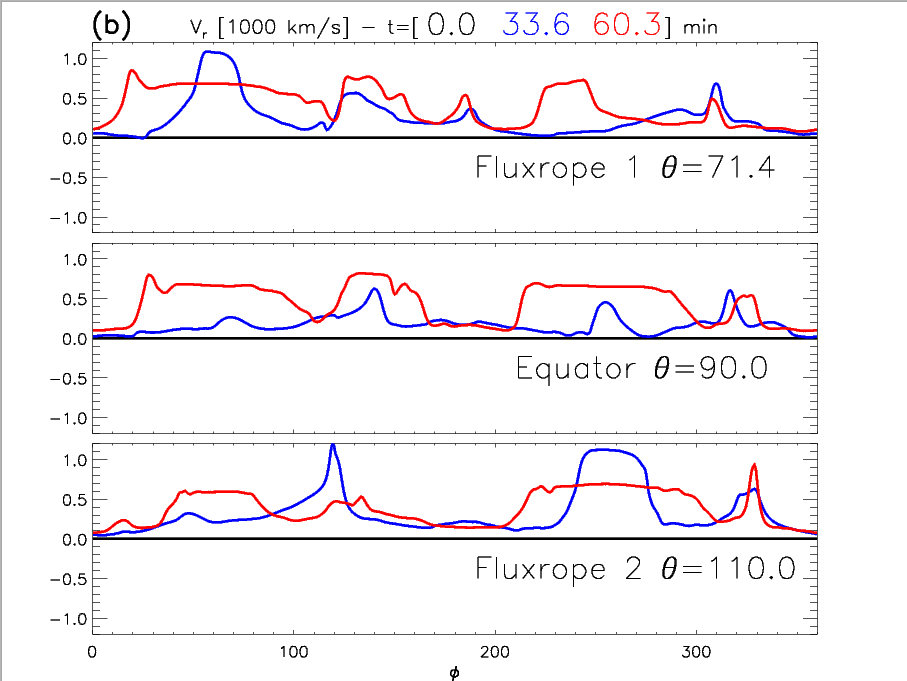}

\includegraphics[scale=0.38,clip]{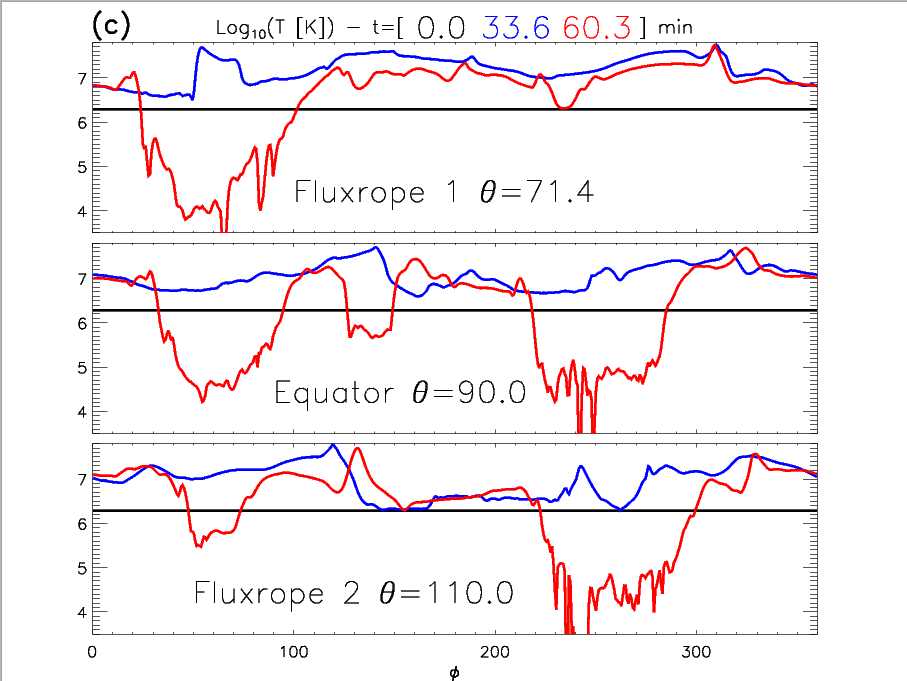}

\caption{Cuts of (a) $Log_{10}(\rho [g/cm^3])$,
(b) radial velocity [cm/s], and (c) temperature [K]
along $\phi$
at $t=0$ $min$ (black line),
$t=33.6$ $min$ (blue line)
and $t=60.3$ $min$ (red line)
at three different co-latitudes:
the center of FR1 ($\theta=71.4^{\circ}$),
the equator ($\theta=90^{\circ}$),
and the center of FR2 ($\theta=110^{\circ}$.)}
\label{outplasmacuts}
\end{figure}
Fig.\ref{outplasmacuts} shows cuts along the $\phi$ direction at the outer boundary at three different co-latitudes
for (a) density, (b) radial velocity and (c) temperature. In each of the plots different co-latitudes
are considered corresponding to
the center of FR1 ($\theta=71.5^{\circ}$, top),  the equator ($\theta=90^{\circ}$, middle)
and finally the center of FR2 ($\theta=110^{\circ}$, bottom). In each plot values are shown at three 
different times: $t=0$ (black line), $t=33.6$ $min$ (blue line) and  $t=60.3$ $min$ (red line). 
The plots allow us to show more quantitatively the inhomogeneous nature at the outer boundary
due to the two ejections.
The latter two times correspond approximately to the times when
the fronts of the ejections reach the outer boundary and when the centres of each flux rope cross it.
It is only approximate as the two fronts and the two centres cross the boundary at slightly different times.
The plots of (a) density and (b) radial velocity both show that the perturbation 
of the quiet corona increases in amplitude as the flux ropes reach and cross the outer boundary.
When the front of each ejection first reaches the outer boundary ($t=33.6$ $min$)
the density increases by less than an order of magnitude and is localised in space. However once the
flux rope crosses the surface ($t=60.3$ $min$) the density
increases by two orders of magnitude. In contrast,  the  radial velocity exhibits a number of different
features. When the fronts from both ejections reach the outer boundary a peak of $\sim1000$ $km/s$ occurs,
after which the radial velocity settles to a more plateau structure of around $500$ $km/s$.
Variations in both the density and velocity  occur over a wide angular degree. Even though 
both of the initial ejections
occur far from the equator, their spatial extension at the outer boundary is wide enough to 
cross into the opposite 
hemisphere. From the cuts taken at the latitudes of either FR1 or FR2 there are clear indications of the
occurrence of a CME in the
opposite hemisphere. In contrast the equatorial cut shows a similar contribution from both
FR1 ($\phi\sim50^{\circ}$ in Fig.\ref{outplasmacuts}) and
FR2 ($\phi\sim250^{\circ}$) in Fig.\ref{outplasmacuts}),
as well as from the polarity inversion line plasma flows ($\phi\sim140^{\circ}$).
Finally the temperature of the plasma crossing the outer boundary qualitatively changes, as first the
front crosses the equator and then the flux rope core. As the fronts from both ejections cross 
the boundary, a temperature increase
is found. However as the flux rope cores pass through the boundary this changes to a temperature
decrease. This behavior seems reasonable from a qualitative point of view, as we expect to find higher
temperatures at the front due to the compression and heating of the plasma and lower temperatures at the 
flux ropes core due to the cooler denser material that is present within it. While we see these 
temperature features we
should also acknowledge that the present model is not designed to model accurately the temperature evolution
as it lacks crucial terms such as thermal conduction and radiative losses in Eq.\ref{energy}.
Even though these terms are not included, they are not
essential for the modelling of the thermal structure of the front and core of the flux rope ejections.

\begin{figure}
\includegraphics[scale=0.13,clip]{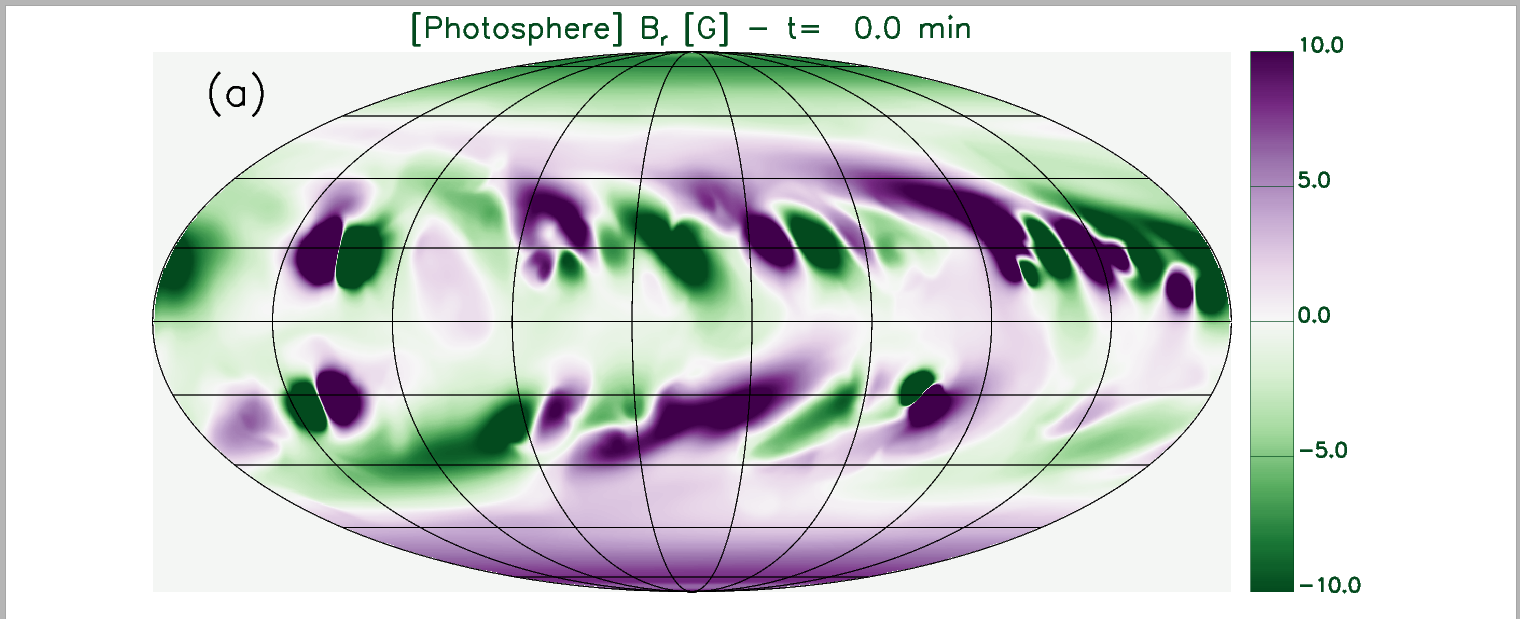}
\includegraphics[scale=0.13,clip]{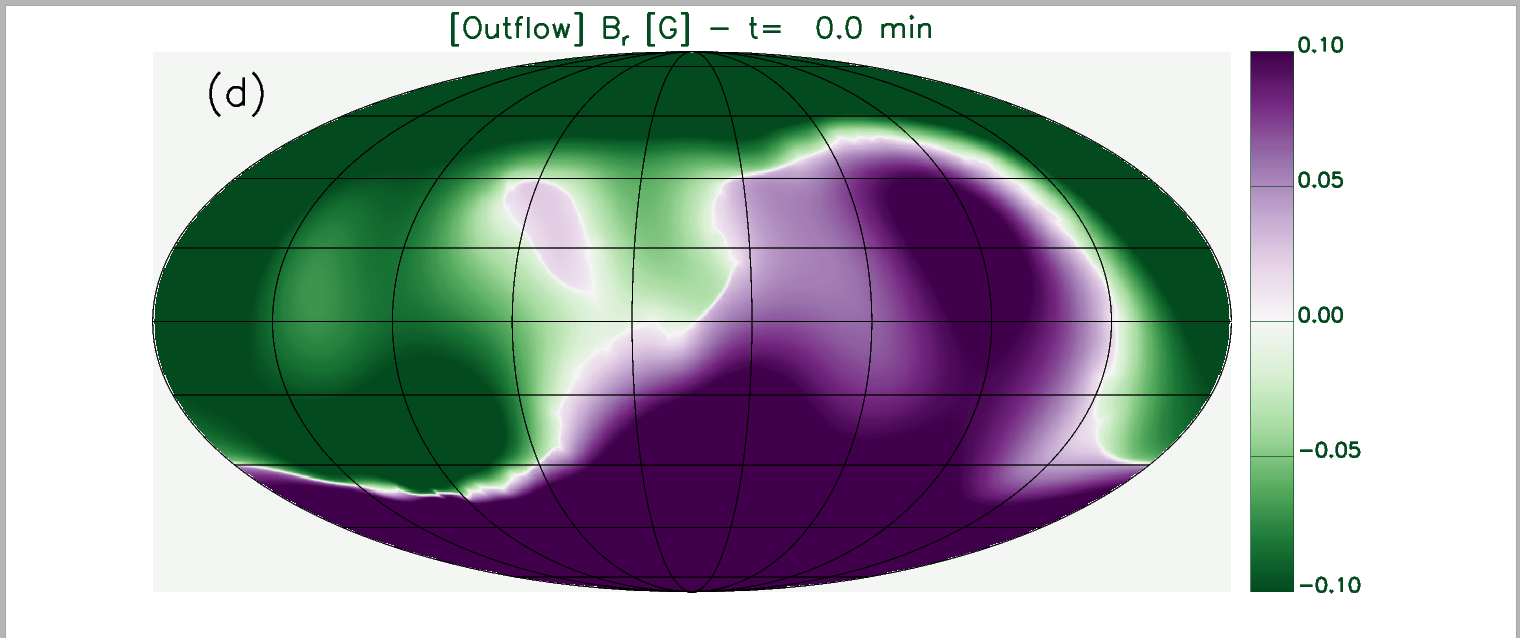}
\includegraphics[scale=0.13,clip]{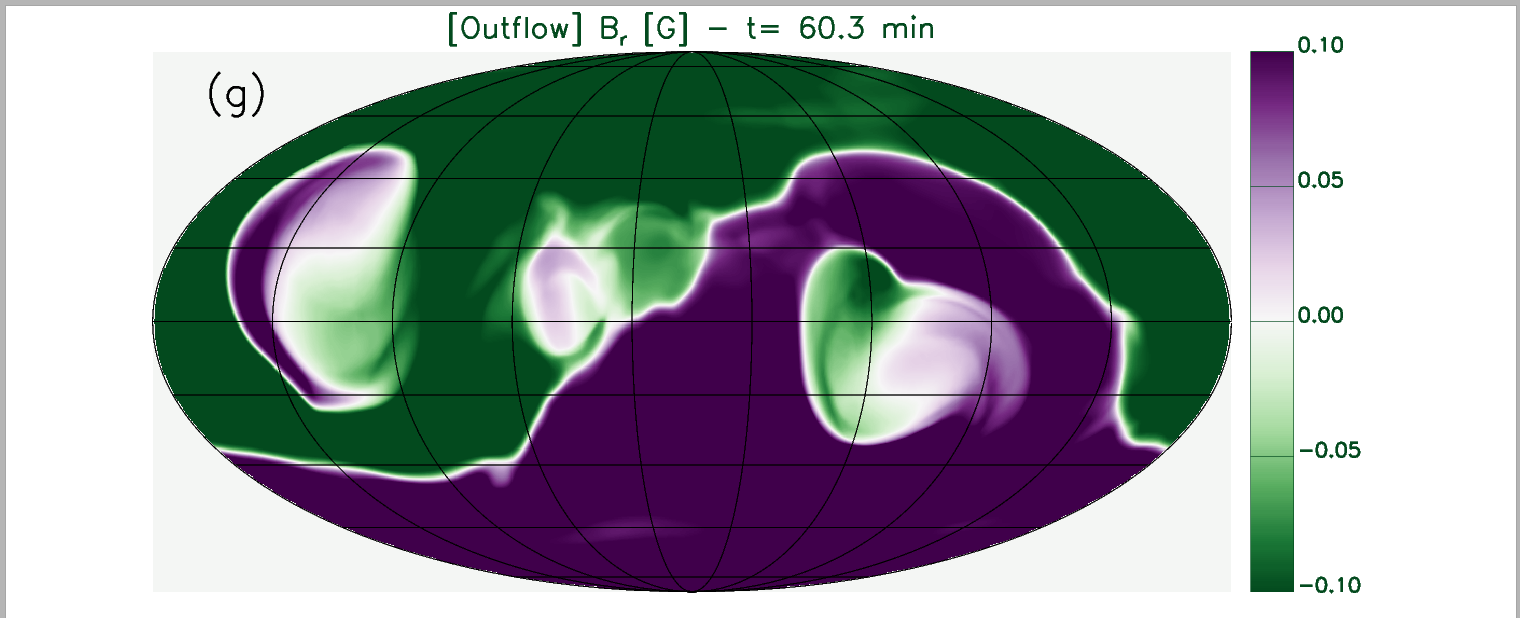}

\includegraphics[scale=0.13,clip]{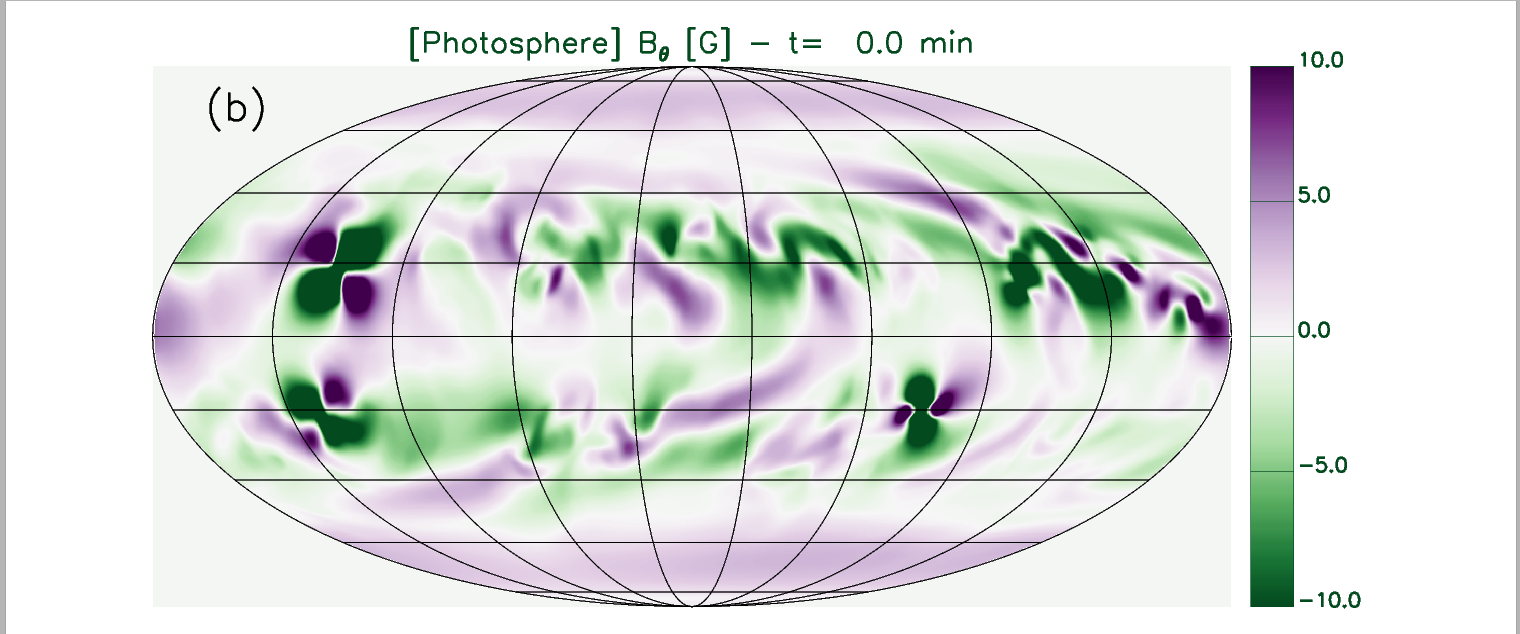}
\includegraphics[scale=0.13,clip]{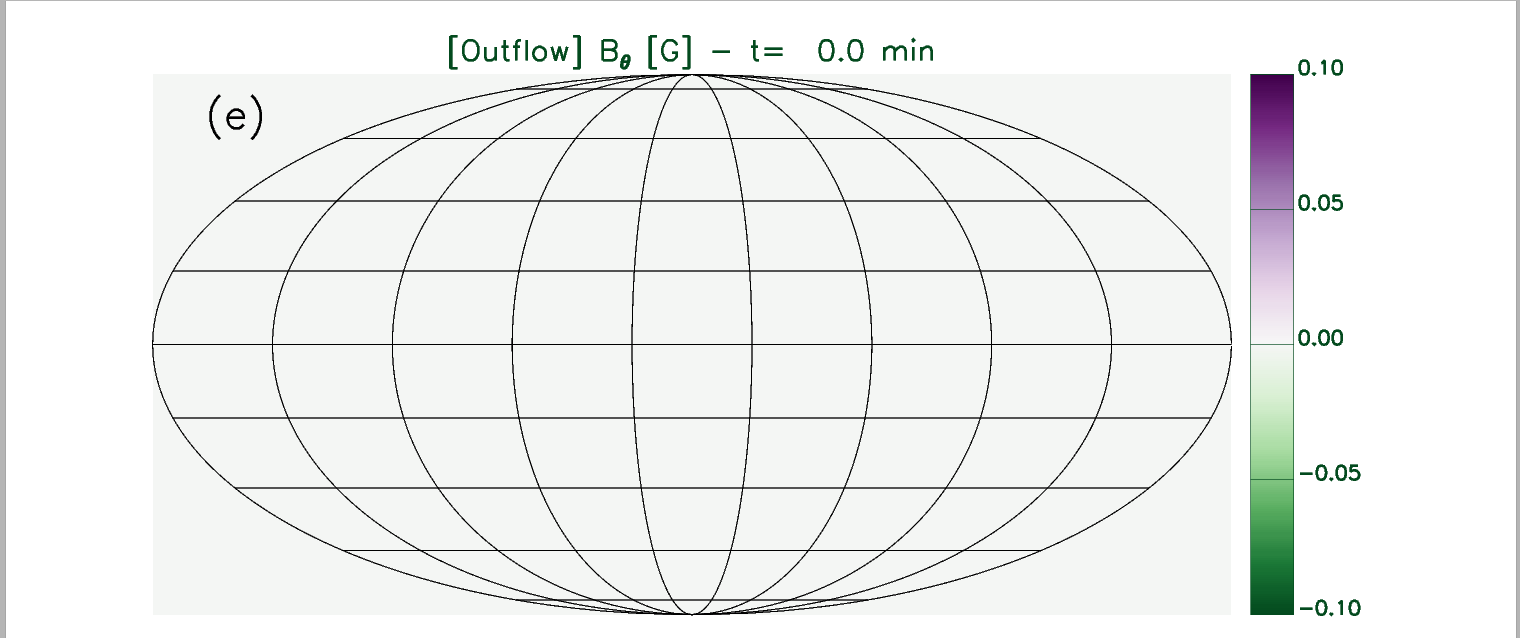}
\includegraphics[scale=0.13,clip]{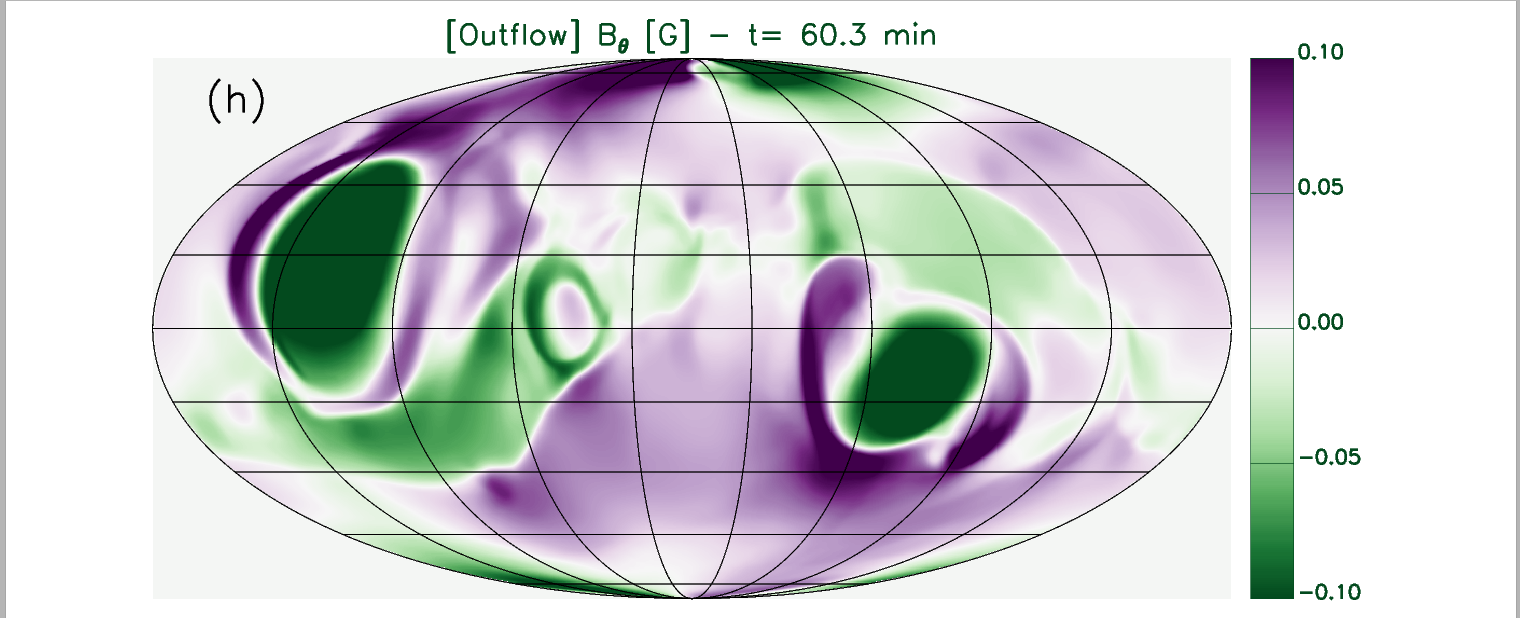}

\includegraphics[scale=0.13,clip]{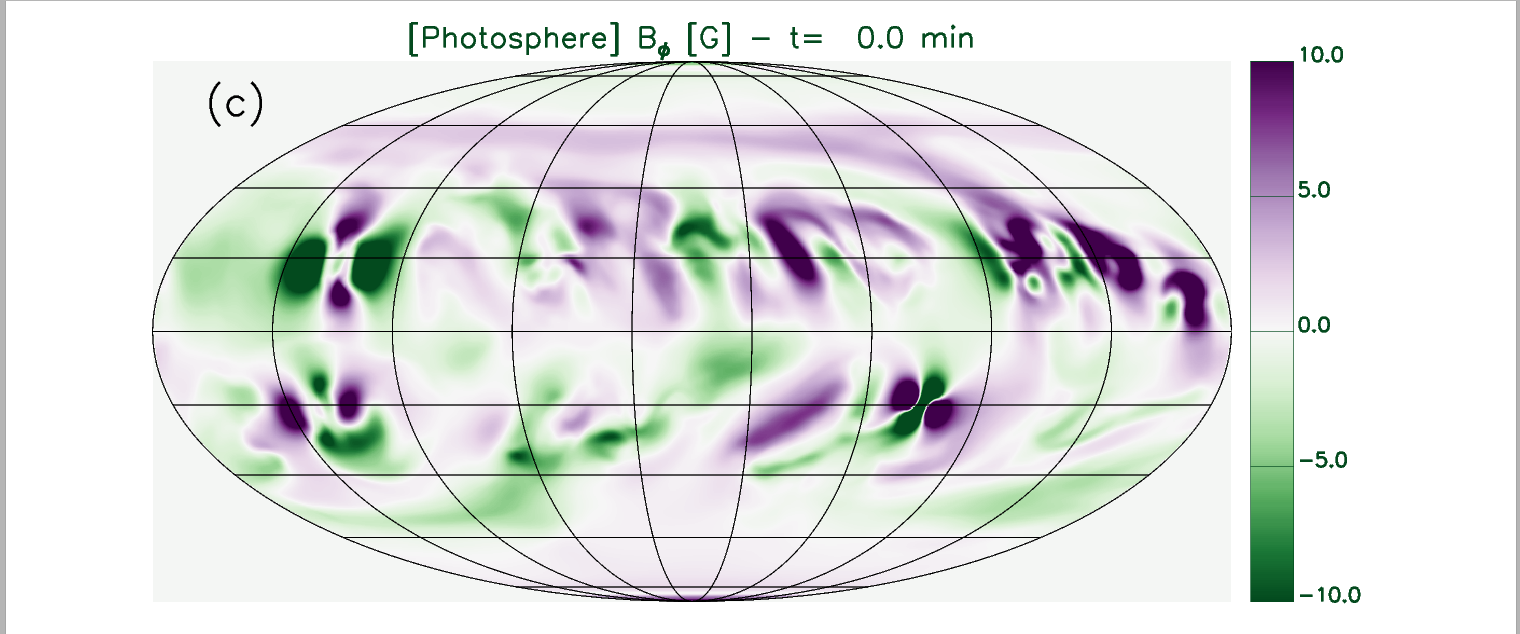}
\includegraphics[scale=0.13,clip]{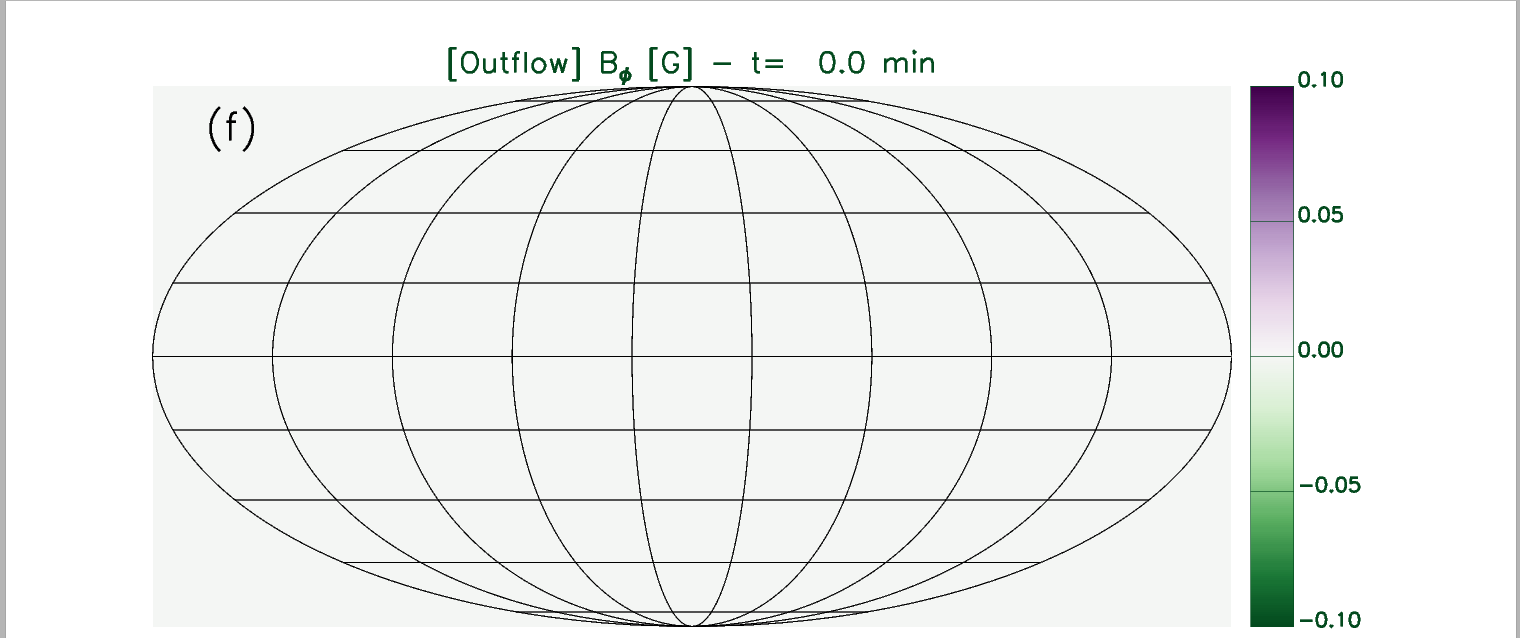}
\includegraphics[scale=0.13,clip]{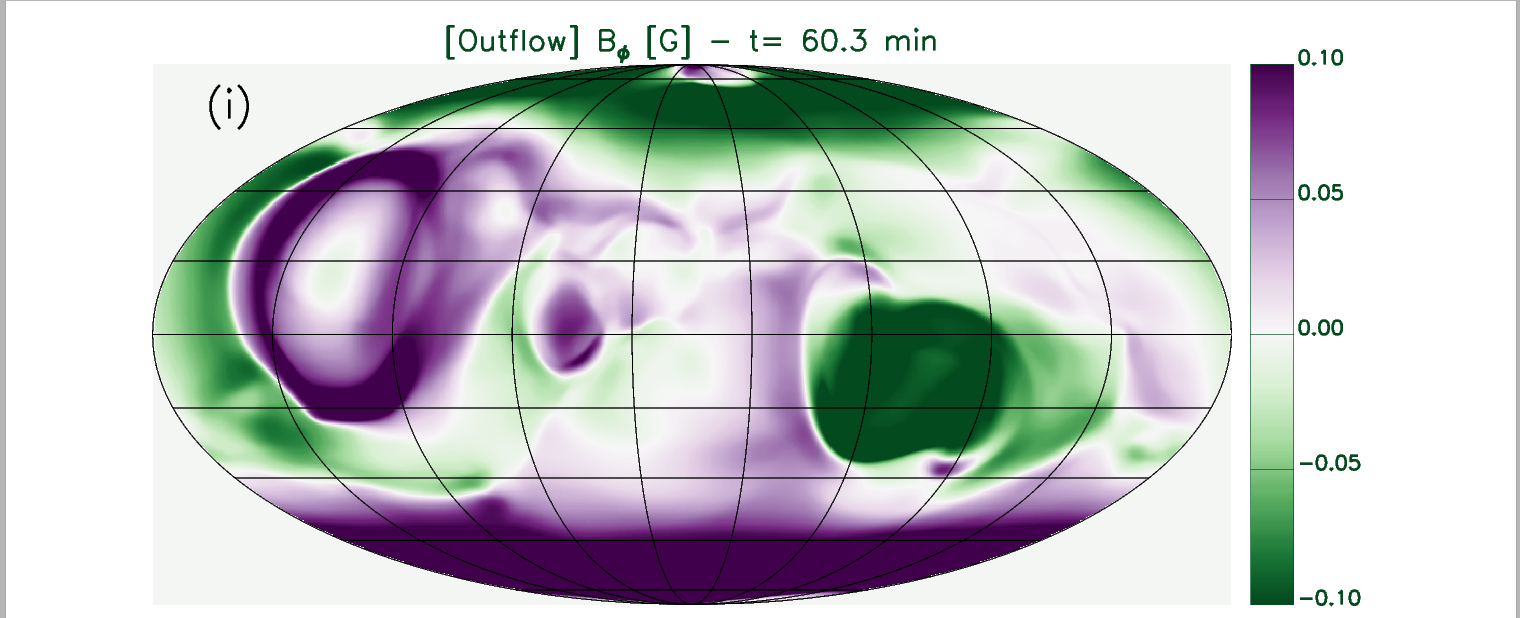}

\caption{Mollweide projections (central meridian at longitude $\phi=180^{\circ}$)
at the lower boundary of the MHD simulation at $t=0$ $min$ (left column)
and at the outer boundary at $t=0$ $min$ (central column) and
$t=60.3$ $min$ (right column)
for the three 
magnetic field components $B_r$, $B_{\theta}$, $B_{\phi}$.}
\label{outmagmaps}
\end{figure}

Fig.\ref{outmagmaps} shows Mollweide projections of the three components of the magnetic field for
(a)-(c) the initial condition of the MHD simulation at the level of the photosphere and 
(d)-(f) at the outer boundary, and 
(g)-(i) at the outer boundary at the time when both flux ropes cross the outer surface ($t=60.3$ $min$). 
The first row shows $B_r$, the middle row $B_{\theta}$ and the bottom row $B_\phi$.
Upon comparing the individual field components at the different heights and at the different times,
it is not surprising that the complexity and structure of the magnetic field at the solar surface (Fig.\ref{outmagmaps}a-c)
is not carried out to the outer corona (Fig.\ref{outmagmaps}d-f).
It is however essential to notice that the configuration of the magnetic field at the outer boundary
is the result of the interaction between the outward travelling structures and the background solar corona.
This interaction simplifies the configuration of the magnetic field
and alters the surviving structures in a way that is not predictable a-priori.
In particular while many magnetic 
features are visible in Fig.\ref{outmagmaps}(a) only two distinct ones appear at the outer boundary during the
eruption (Fig.\ref{outmagmaps}(g)).
At the outer boundary the flux rope PILs are of different shape and size, even 
though the two flux ropes are initially of the same size. Finally, both ejections show a large 
negative $B_\theta$ patch at the outer boundary, 
but opposite and different $B_\phi$ patches showing that they undergo a different rotation.

By comparing Fig.\ref{outmagmaps}(g)-(i) with Fig.\ref{outmagmaps}(d)-(f),
it is clear that the magnetic field carried by the 
flux ropes significantly perturbs the outer boundary. The radial component of the magnetic 
field exhibits at $t=0$ $s$ a clearly defined distribution, where there are two regions with opposite polarity that are
separated by a polarity inversion line.
Within these locations when the flux rope ejection
occurs and reaches the outer boundary it alters the magnetic field distribution.
The distribution becomes more in-homogeneous and structured.
Within each region that was originally of a single polarity,
opposite polarity patches now appear indicating the flux rope ejections.
In addition to the changes in the radial field,
as the initial magnetic field is prescribed to be only radial at the outer boundary,
the $\theta$ and $\phi$ components at the outer boundary show a major change
once the erupting flux rope reaches it.
Both erupting flux ropes carry a negative $\theta$ component creating two areas (green) clearly visible in a 
more uniform and less intense background. Finally, a clear signature of the flux ropes can also be seen 
in the $\phi$ component of the magnetic field, where once again two very visible regions are present.
An interesting feature is that around FR1 a distinct ring structure of positive $B_{\phi}$ is present, within which
there is contained a weak $B_{\phi}$. 
In contrast around FR2, there is a more uniform
negative $B_{\phi}$. The origin of this variation is still under investigation, however it illustrates the importance
of modelling realistic photospheric magnetic field configurations both before and  during the eruption as variations
at the outer boundary must ultimately be due to variations in the low coronal field. 
Thus within the present simulation due to the low coronal properties of the field and the regions in
which the flux ropes form, we find significant differences in the field distribution at the outer
boundary. If the low coronal field is not accurately modelled then such a variation in ICME's will not
be taken into account.

\begin{figure}
\includegraphics[scale=0.30,clip]{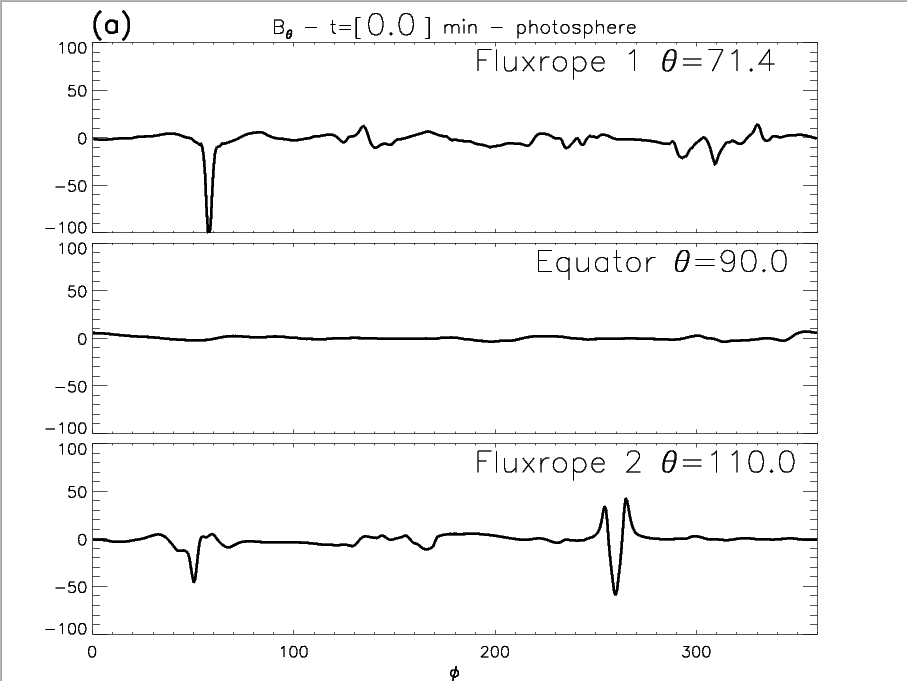}
\includegraphics[scale=0.30,clip]{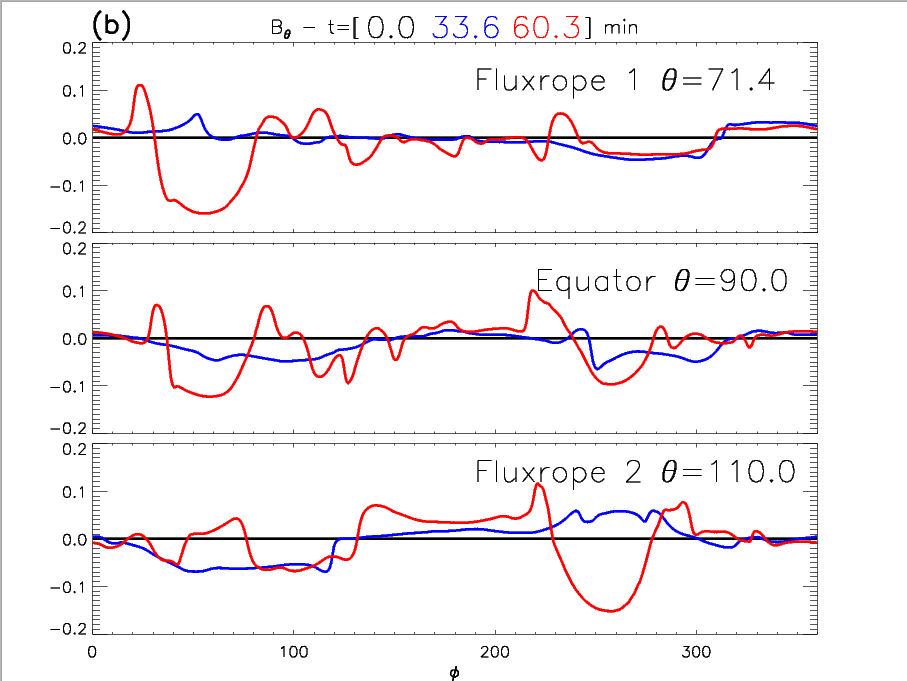}

\caption{Cuts of $B_{\theta}$
along $\phi$
at three different latitudes:
the center of FR1 ($\theta=71.4^{\circ}$),
the equator ($\theta=90^{\circ}$),
and the center of FR2 ($\theta=110^{\circ}$).
a) At $t=0$ and lower boundary.
b) At $t=0$ $min$ (black line),
$t=33.6$ $min$ (blue line)
and $t=60.3$ $min$ (red line) and outer boundary
}
\label{outbzcuts}
\end{figure}
In Fig.\ref{outbzcuts} we focus on the time evolution of $B_\theta$ at the latitudes of FR1 (top), 
the equator (middle)  and FR2 (bottom). Fig.\ref{outbzcuts}(a) shows the quantity at the level of the photosphere
at time $t=0.0$ just before the eruption starts. Fig.\ref{outbzcuts}(b) shows the results for the outer boundary
at the times of $t=0.0$ (black line), $t=33.6$ (blue line) and $t=60.3$ (red line).
Again comparisons between Fig.\ref{outbzcuts}(a) and Fig.\ref{outbzcuts}(b) show no evident correlation
between the $B_\theta$ values at the lower boundary before the onset of ejections to that of the resulting configuration
at the outer boundary as the eruption passes through it. 

In Fig.\ref{outbzcuts}(b) the $\theta$ component of the magnetic field is particularly important as it gives the
out-of-ecliptic plane component at the equator. For space weather consequences the intensity and orientation of
this component when the ICME encounters the Earth magnetosphere determines  the geo-effectiveness of the disturbance.
At the present time it is not clear if the out-of-ecliptic component of the magnetic field  at 4 $R_\odot$ is 
preserved during the transit of the CME to Earth. However in order to produce accurate predictions at Earth 
of its intensity and orientation accurate initial conditions are required when it is ejected into interplanetary space.
In Fig.\ref{outbzcuts}(b) we find that the ejection front generates a modest enhancement in the $B_\theta$
component at all three latitudes. However the overall effect of each front is quite wide.
At the equator when the two fronts reach the outer boundary we find a preference for 
negative $B_\theta$ at the longitudes corresponding to the flux ropes.
A more significant $B_\theta$ enhancement occurs when the flux ropes themselves reach the outer boundary.
This is most
intense at the center of the flux rope locations, but still visible at other longitudes.
At the equator, we find a strong negative $B_\theta$ value at the longitudes of the two flux ropes,
with a positive peak between them.

\begin{figure}
\includegraphics[scale=0.30,clip]{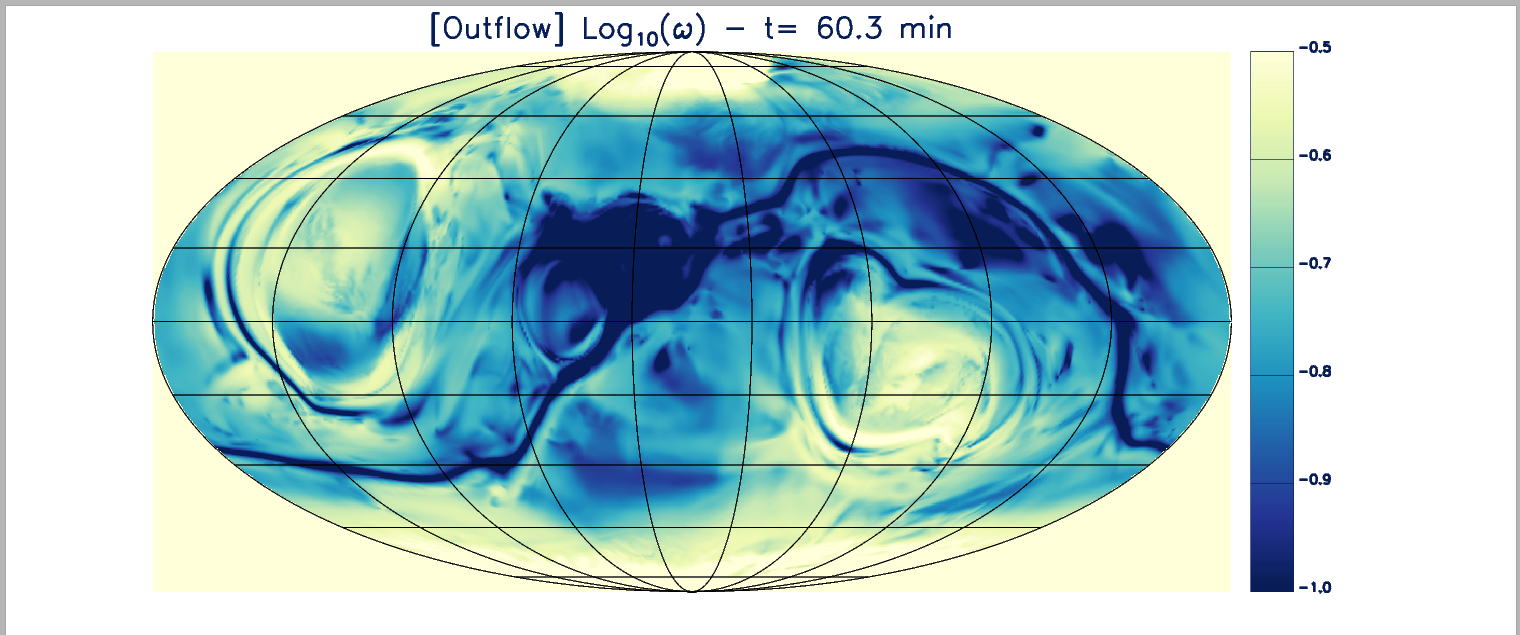}
\caption{Mollweide projection of $\omega$ at the outer boundary of the MHD simulation at $t=60.3$ $min$
(central meridian at longitude $\phi=180^{\circ}$) .}
\label{outomegamap}
\end{figure}
Finally in Fig.\ref{outomegamap} we show a Mollweide projection of $\omega$ at the outer boundary at $t=60.3$ $min$ when the flux rope 
crosses the surface. This is to determine whether or not the function $\omega$
is appropriate to identify the flux ropes at higher radial distances.
Analysis indicates that the locations involved with the flux rope ejections show up as visible features in 
the  $\omega$ distribution. These generally occur at high $\omega$ values even if it is not possible to 
define a simple 
threshold that isolates the flux rope. It is also clear from Fig.\ref{outomegamap} that the polarity inversion 
line is also a very clear feature due to the relatively lower intensity and less twisted configuration.

\section{Discussion, Outlook, and Conclusions}
\label{conclusion}
The work presented here is the first step of a longer-term effort to improve our predictive capability
for the occurrence of events on the Sun relevant to space weather. As part of this process we will also improve
our understanding of the evolution of non-potential magnetic structures in the solar corona
and how they lead to CMEs.

\subsection{Present Results}
In this paper we have carried out global simulations of the Sun where the entire extent of the 
corona is considered from the photosphere at $R_{\odot}$ out to $4R_\odot$.
Two models have been coupled to follow the full life cycle of flux
ropes from formation to eruption. Initially, a quasi-static non-potential model is used to consider the slow evolution
of the coronal magnetic field and formation of flux ropes up until the point of eruption. To follow the dynamics of the eruption
an MHD simulation is then carried out. As the quasi-static non-potential simulation does not provide the MHD simulation
with a density profile, we put forward a new technique to prescribe the density using the twisted nature of the
magnetic field. This allows us to produce an inhomogeneous corona, along with flux ropes that have a density 1-2 orders
of magnitude greater than that of the surrounding corona.
The simulation presented in this paper shows the ejection of two separate magnetic flux ropes
that have formed in the global solar corona at different locations.
These flux ropes have been formed self-consistently in magnetic field configurations determined directly from observations.
In the simulations the onset of eruption occurs once the tension of the 
overlying arcades is insufficient to hold down the underlying flux rope.
This occurs depending on the size of the flux rope and the strength/topology
of the overlying arcades and represents the occurrence of a non-equilibrium. When
a non-equilibrium occurs and the flux rope starts to rise, reconnection then occurs 
under it which results in a radially outward Lorentz force which ejects the flux rope. 
This can be seen as a two step process of initial slow rise, followed by a faster ejection 
\citep[see][for a description]{MackayVanBallegooijen2006A,MackayVanBallegooijen2006B}.

In the present model the flux ropes are ejected out of the computational domain as
the Lorentz forces generated in the corona due to the surface magnetic reconnection and shear are too large to be 
contained by the overlying arcades \citep{Yeates2009}. 
During the ejections, both flux ropes show velocity 
values along with magnetic and density structures compatible with standard CME parameters at $4$ $R_{\odot}$.

When the ejections reach the outer boundary at $4$ $R_{\odot}$ we discuss how the CME passes over this boundary and 
is injected into the solar wind. We find that the flux rope ejection into the solar wind occurs over an area 
of tenths of square degree, where the ejected plasma is denser compared to that of the quiet corona. In addition the
ejected plasma and magnetic field show a highly inhomogeneous structure.
We highlight here the importance, for this study, of the approximation we have made to 
construct a realistic atmosphere around a magnetic configuration that contains flux ropes (given by
Eq. \ref{omegab},\ref{omegarthetaphi} and \ref{tempomega}). This allows us to account for the 
diversity of density in the coronal environment taking into account the importance of the magnetic field in 
shaping the structures in the solar corona, including dense horizontal structures.

An essential result from our modelling work is that the propagation of the ejection and 
the consequent perturbation at $4$ $R_{\odot}$ is highly sensitive to the local pre-eruption conditions
as the two analysed ejections carry different kinematic and plasma properties. 
Both also show no simple connection between the flux rope configuration at the lower boundary and 
its following evolution.
Due to this, each flux rope ejection should be considered a unique phenomenon and
consequently its associated space weather perturbation will also be unique.

In the present paper we have shown the potential capability of this technique through a specific example.
An essential follow up to this work will be to compare models of magnetic flux rope ejections with observations.
The quasi-static non-potential global model has been thoroughly tested for the identification of actual magnetic flux rope formation 
and times of eruptions.
In the future we will also compare the magnetic flux rope evolution as described by the MHD simulation
with observations of specific events.

\subsection{Producing an Operational Model}
The present model shows many interesting features that with further development may be a useful  
tool for application to space weather prediction.
The technique presented could provide realistic time dependent inhomogeneous boundary conditions that couple the
outer corona to the heliosphere. This coupling still requires development and testing within a heliospheric model. 
In particular, we have shown that from a single magnetic configuration we may
produce multiple eruptions, where each of these eruptions have an inhomogeneous structure to the magnetic field
and density as  they pass through the interface between the corona and interplanetary space. Another interesting outcome is that
the magnetic field and density properties of the two eruptions have significant differences in size and distribution.
Therefore to some degree each eruption is unique, based on the initial properties of the flux rope and the 
surrounding non-potential field produced in the quasi-static evolution model.

For practical applications any space weather forecasting tool needs to have two key qualities: accuracy and 
efficiency. Accuracy is required in order to avoid false warnings and to ensure key parameters such as the
arrival time of the CME at Earth and the orientation of this magnetic field and density are reliably determined.
In terms of this our model sets high standards of accuracy as firstly it evolves the global corona from magnetic 
configurations determined from the time evolution of observed magnetograms. This produces non-potential magnetic 
configurations very close to observations and 
the global non-potential coronal model has proved highly reliable in predicting the locations, helicity and timing
of flux rope formations, especially after the model has taken into account the emergence of new magnetic flux.
Secondly, it solves the full MHD equations to follow the dynamical eruption itself. 
For time critical predictions efficiency is essential to run useful simulations that can 
output results before or close to when an event has taken place. In our tests, the global non-potential model is more than 
$10^{4}$ times computationally faster than a full MHD simulation. This is mostly due to the lower number of equations 
to solve and to the higher time steps involved when Alfv\'en waves do not need to be resolved. If a single MHD model was used, then
the vast majority of the physical time of the simulation would be during the formation of the flux rope (days to weeks)
compared to hours for the eruption. Thus use of the global non-potential model to produce the initial eruptive
configuration is twofold, first it solves fewer equations and secondly it is much more efficient. This means that only
a small part of the physical time (once the ejection sets in) needs to be considered and modeled with a full MHD
simulation. The combined  model is in principle capable of predicting flux rope formation and subsequent 
onset of ejection days before it occurs, where the MHD simulation would only need to simulate a few hours of physical time
which is well within current computational capacity. While this is a positive aspect of this approach, before such a
technique can become viable as a semi-autonomous, factually accurate,  operational space weather forecasting model,
a number of practical and scientific obstacles need to be overcome. These include
\begin{enumerate}
\item {\bf Flux emergence accuracy.\\} The current global non-potential  model is highly dependent on a semi-automated determination 
of flux emergence from observations. This is key as the emergence of new flux can significantly alter the magnetic 
configuration of the solar corona. Therefore a more accurate treatment must be developed. Techniques are under development 
and testing that allow
for the  direct assimilation of normal component magnetograms across all longitudes and latitudes
(see \cite{Weinzierl2016a} and \cite{Weinzierl2016b}).
One issue with the direct assimilation of magnetogram data into the simulation, is that only a portion of the global photospheric field
may be observed at any given time. However, if magnetogram data were also available from an L5 based magnetograph this would significantly
enhance the accuracy of the models. In the recent paper by \cite{Mackay2016} it has been shown that such L5 observations could 
increase the accuracy
of global quantities relevant to eruptions on the Sun by anywhere from 26-40$\%$.
\citet{Pevtsov2016} reports similar improvements for the estimations of solar wind and plasma outflows as well.
\item {\bf Identification of erupting flux ropes.\\}
Although the global non-potential model is able to describe the formation of magnetic flux ropes, it cannot account for the 
whole physics of an eruption. For the latter MHD simulations are employed. Therefore during the global non-potential
simulations automated quantitative analysis of the properties of the coronal magnetic field must be carried out to first identify
the existence of flux ropes and then the onset of ejections. A number of candidates for this task include instability analysis, 
measurement of thresholds, or recognizing critical indices in the parameters of the magnetic field configuration.
\item {\bf Coupling AMRVAC back to the Global Non-potential Model.\\}
One of the main strengths of the global non-potential modelling technique is the continuous nature of the simulations
over days to years, in which the transport of magnetic flux and magnetic helicity from low to high latitudes is followed. 
Once magnetic field configuration are imported from the global non-potential model into ARMVAC the continunity of 
the global non-potential simulations ends. Thus in order to allow a continuous modelling of the solar coronal evolution 
after the eruption,
it is vital to couple the related MHD solution back to the global non-potential model. This will then be used to
start a new segment of the quasi-static evolution. Currently this is not possible due to technical limitations 
where the primary variable in the  Global Non-potential Model is the  vector potential $\vec{A}$, while in 
MPI-AMRVAC it is the  magnetic field $\vec{B}$. To resolve this issue a new MPI-AMRVAC module that solves the MHD 
equations using $\vec{A}$ is under development.
\item {\bf Matching the simulations with the lower boundary of space weather models.\\}
Most space weather forecast tools focus on the physical domain extending from $0.1 AU$ to $>1$ $AU$.
The lower boundary of this range corresponds to about $21$ $R_{\odot}$
and is significantly higher than the outer boundary of the simulation shown here.
To bridge the gap between the models
a number of options will be investigated in the future.
These include using a simple solar wind model above $4$ $R_{\odot}$ \citep[for a review:][]{Echim2011}
or to scale the plasma and velocity fluxes from $4$ $R_{\odot}$ up to $21$ $R_{\odot}$.
In the latter approach we would consider it as a perturbation to a background solar wind that is included
in the applied space weather forecast tool.
In future studies we will consider the validity and technical aspects of both of these approaches.
\end{enumerate}

In conclusion, the present work shows that in the future the coupling of the Global Model with MHD simulations
may become a useful and essential tool to bridge magnetogram observations
with space weather modellling. This is due to the complexity of the magnetic configurations generated in the solar corona
and the interaction between propagating flux ropes and the coronal environment.
The technique we put forward is a viable one for future space weather forecasting tools and
has the potential not only to improve our forecasting capabilities, but also to lead to a better understanding
of the physics of CME origin.

\begin{acknowledgements}
This research has received funding from the European Research Council (ERC) under the European Union's Horizon 2020 research and innovation programme (grant agreement No 647214).
This work used the DiRAC Data Centric system at Durham University, operated by the Institute for Computational Cosmology on behalf of the STFC DiRAC HPC Facility (www.dirac.ac.uk. This equipment was funded by a BIS National E-infrastructure capital grant ST/K00042X/1, STFC capital grant ST/K00087X/1, DiRAC Operations grant ST/K003267/1 and Durham University. DiRAC is part of the National E-Infrastructure.
We acknowledge the use of the open source (gitorious.org/amrvac) MPI-AMRVAC software, relying on coding efforts from C. Xia, O. Porth, R. Keppens.
D.H.M. would like to thank STFC and the Leverhulme Trust for their financial support.
ARY was supported by STFC consortium grant  ST/N000781/1 to the universities of Dundee and Durham.
The editor thanks two anonymous referees for their assistance in evaluating this paper.
\end{acknowledgements}

\bibliography{ref}

\begin{thebibliography}{65}
\providecommand{\natexlab}[1]{#1}
\providecommand{\url}[1]{\texttt{#1}}
\providecommand{\urlprefix}{URL }
\providecommand{\eprint}[2][]{\url{#2}}

\bibitem[{{Archontis} and {Hood}(2012)}]{ArchontisHood2012}
{Archontis}, V., and A.~W. {Hood}.
\newblock {Magnetic flux emergence: a precursor of solar plasma expulsion}.
\newblock \emph{\aap}, \textbf{537}, A62, 2012.
\newblock 10.1051/0004-6361/201116956.

\bibitem[{{Baker} et~al.(2013){Baker}, {Poh}, {Odstrcil}, {Arge}, {Benna}
  et~al.}]{Baker2013}
{Baker}, D.~N., G.~{Poh}, D.~{Odstrcil}, C.~N. {Arge}, M.~{Benna}, et~al.
\newblock {Solar wind forcing at Mercury: WSA-ENLIL model results}.
\newblock \emph{Journal of Geophysical Research (Space Physics)}, \textbf{118},
  45--57, 2013.
\newblock 10.1029/2012JA018064.

\bibitem[{{Bisi} et~al.(2013){Bisi}, {Jackson}, {Fallows}, {Tokumaru},
  {Jensen}, {Lee}, {Harrison}, {Hapgood}, {Wu}, and {Davies}}]{Bisi2013}
{Bisi}, M.~M., B.~V. {Jackson}, R.~A. {Fallows}, M.~{Tokumaru}, E.~A. {Jensen},
  J.~{Lee}, R.~{Harrison}, M.~A. {Hapgood}, C.~{Wu}, and J.~{Davies}.
\newblock {Using Interplanetary Scintillation (IPS) For Space-Weather
  Forecasting}.
\newblock \emph{AGU Fall Meeting Abstracts}, 2013.

\bibitem[{{Cheng} et~al.(2011){Cheng}, {Zhang}, {Liu}, and {Ding}}]{Cheng2011}
{Cheng}, X., J.~{Zhang}, Y.~{Liu}, and M.~D. {Ding}.
\newblock {Observing Flux Rope Formation During the Impulsive Phase of a Solar
  Eruption}.
\newblock \emph{\apjl}, \textbf{732}, L25, 2011.
\newblock 10.1088/2041-8205/732/2/L25, \eprint{1103.5084}.

\bibitem[{{Chintzoglou} et~al.(2015){Chintzoglou}, {Patsourakos}, and
  {Vourlidas}}]{Chintzoglou2015}
{Chintzoglou}, G., S.~{Patsourakos}, and A.~{Vourlidas}.
\newblock {Formation of Magnetic Flux Ropes during a Confined Flaring Well
  before the Onset of a Pair of Major Coronal Mass Ejections}.
\newblock \emph{\apj}, \textbf{809}, 34, 2015.
\newblock 10.1088/0004-637X/809/1/34, \eprint{1507.01165}.

\bibitem[{{Dewey} et~al.(2015){Dewey}, {Baker}, {Anderson}, {Benna}, {Johnson}
  et~al.}]{Dewey2015}
{Dewey}, R.~M., D.~N. {Baker}, B.~J. {Anderson}, M.~{Benna}, C.~L. {Johnson},
  et~al.
\newblock {Improving solar wind modeling at Mercury: Incorporating transient
  solar phenomena into the WSA-ENLIL model with the Cone extension}.
\newblock \emph{Journal of Geophysical Research (Space Physics)}, \textbf{120},
  5667--5685, 2015.
\newblock 10.1002/2015JA021194.

\bibitem[{{D'Huys} et~al.(2014){D'Huys}, {Seaton}, {Poedts}, and
  {Berghmans}}]{DHuys2014}
{D'Huys}, E., D.~B. {Seaton}, S.~{Poedts}, and D.~{Berghmans}.
\newblock {Observational Characteristics of Coronal Mass Ejections without
  Low-coronal Signatures}.
\newblock \emph{\apj}, \textbf{795}, 49, 2014.
\newblock 10.1088/0004-637X/795/1/49, \eprint{1409.1422}.

\bibitem[{{Echim} et~al.(2011){Echim}, {Lemaire}, and
  {Lie-Svendsen}}]{Echim2011}
{Echim}, M.~M., J.~{Lemaire}, and {\O}.~{Lie-Svendsen}.
\newblock {A Review on Solar Wind Modeling: Kinetic and Fluid Aspects}.
\newblock \emph{Surveys in Geophysics}, \textbf{32}, 1--70, 2011.
\newblock 10.1007/s10712-010-9106-y, \eprint{1306.0704}.

\bibitem[{{Falkenberg} et~al.(2011){Falkenberg}, {Taktakishvili}, {Pulkkinen},
  {Vennerstrom}, {Odstrcil}, {Brain}, {Delory}, and
  {Mitchell}}]{Falkenberg2011}
{Falkenberg}, T.~V., A.~{Taktakishvili}, A.~{Pulkkinen}, S.~{Vennerstrom},
  D.~{Odstrcil}, D.~{Brain}, G.~{Delory}, and D.~{Mitchell}.
\newblock {Evaluating predictions of ICME arrival at Earth and Mars}.
\newblock \emph{Space Weather}, \textbf{9}, S00E12, 2011.
\newblock 10.1029/2011SW000682.

\bibitem[{{Gopalswamy} et~al.(2000){Gopalswamy}, {Lara}, {Lepping}, {Kaiser},
  {Berdichevsky}, and {St.~Cyr}}]{Gopalswamy2000b}
{Gopalswamy}, N., A.~{Lara}, R.~P. {Lepping}, M.~L. {Kaiser},
  D.~{Berdichevsky}, and O.~C. {St.~Cyr}.
\newblock {Interplanetary Acceleration of Coronal Mass Ejections}.
\newblock \emph{\grl}, \textbf{27}, 145--148, 2000.

\bibitem[{{Gopalswamy} et~al.(2003){Gopalswamy}, {Shimojo}, {Lu}, {Yashiro},
  {Shibasaki}, and {Howard}}]{Gopalswamy2003b}
{Gopalswamy}, N., M.~{Shimojo}, W.~{Lu}, S.~{Yashiro}, K.~{Shibasaki}, and
  R.~A. {Howard}.
\newblock {Prominence Eruptions and Coronal Mass Ejection: A Statistical Study
  Using Microwave Observations}.
\newblock \emph{\apj}, \textbf{586}, 562--578, 2003.
\newblock 10.1086/367614.

\bibitem[{{Hapgood}(2011)}]{Hapgood2011}
{Hapgood}, M.~A.
\newblock {Towards a scientific understanding of the risk from extreme space
  weather}.
\newblock \emph{Advances in Space Research}, \textbf{47}, 2059--2072, 2011.
\newblock 10.1016/j.asr.2010.02.007.

\bibitem[{{Harrison} et~al.(2017){Harrison}, {Davies}, {Biesecker}, and
  {Gibbs}}]{Harrison2017}
{Harrison}, R.~A., J.~A. {Davies}, D.~{Biesecker}, and M.~{Gibbs}.
\newblock {The application of heliospheric imaging to space weather operations:
  Lessons learned from published studies}.
\newblock \emph{Space Weather}, \textbf{15}, 985--1003, 2017.
\newblock 10.1002/2017SW001633.

\bibitem[{{Howard} and {DeForest}(2014)}]{Howard2014}
{Howard}, T.~A., and C.~E. {DeForest}.
\newblock {The Formation and Launch of a Coronal Mass Ejection Flux Rope: A
  Narrative Based on Observations}.
\newblock \emph{\apj}, \textbf{796}, 33, 2014.
\newblock 10.1088/0004-637X/796/1/33.

\bibitem[{{Howard} and {Tappin}(2009)}]{TappinHoward2009a}
{Howard}, T.~A., and S.~J. {Tappin}.
\newblock {Interplanetary Coronal Mass Ejections Observed in the Heliosphere:
  1. Review of Theory}.
\newblock \emph{\ssr}, \textbf{147}, 31--54, 2009.
\newblock 10.1007/s11214-009-9542-5.

\bibitem[{{Hutton} and {Morgan}(2015)}]{Hutton2015}
{Hutton}, J., and H.~{Morgan}.
\newblock {Erupting Filaments with Large Enclosing Flux Tubes as Sources of
  High-mass Three-part CMEs, and Erupting Filaments in the Absence of Enclosing
  Flux Tubes as Sources of Low-mass Unstructured CMEs}.
\newblock \emph{\apj}, \textbf{813}, 35, 2015.
\newblock 10.1088/0004-637X/813/1/35.

\bibitem[{{Jackson} et~al.(2010){Jackson}, {Hick}, {Bisi}, {Clover}, and
  {Buffington}}]{Jackson2010}
{Jackson}, B.~V., P.~P. {Hick}, M.~M. {Bisi}, J.~M. {Clover}, and
  A.~{Buffington}.
\newblock {Inclusion of In-Situ Velocity Measurements into the UCSD
  Time-Dependent Tomography to Constrain and Better-Forecast Remote-Sensing
  Observations}.
\newblock \emph{\solphys}, \textbf{265}, 245--256, 2010.
\newblock 10.1007/s11207-010-9529-0.

\bibitem[{{Jin} et~al.(2017){Jin}, {Manchester}, {van der Holst}, {Sokolov},
  {T{\'o}th}, {Mullinix}, {Taktakishvili}, {Chulaki}, and {Gombosi}}]{Jin2017}
{Jin}, M., W.~B. {Manchester}, B.~{van der Holst}, I.~{Sokolov}, G.~{T{\'o}th},
  R.~E. {Mullinix}, A.~{Taktakishvili}, A.~{Chulaki}, and T.~I. {Gombosi}.
\newblock {Data-constrained Coronal Mass Ejections in a Global
  Magnetohydrodynamics Model}.
\newblock \emph{\apj}, \textbf{834}, 173, 2017.
\newblock 10.3847/1538-4357/834/2/173, \eprint{1605.05360}.

\bibitem[{{Li} and {Zhang}(2013)}]{LiZhang2013}
{Li}, L.~P., and J.~{Zhang}.
\newblock {Eruptions of two flux ropes observed by SDO and STEREO}.
\newblock \emph{\aap}, \textbf{552}, L11, 2013.
\newblock 10.1051/0004-6361/201221005.

\bibitem[{{Mackay} and {van
  Ballegooijen}(2006{\natexlab{a}})}]{MackayVanBallegooijen2006A}
{Mackay}, D.~H., and A.~A. {van Ballegooijen}.
\newblock {Models of the Large-Scale Corona. I. Formation, Evolution, and
  Liftoff of Magnetic Flux Ropes}.
\newblock \emph{\apj}, \textbf{641}, 577--589, 2006{\natexlab{a}}.
\newblock 10.1086/500425.

\bibitem[{{Mackay} and {van
  Ballegooijen}(2006{\natexlab{b}})}]{MackayVanBallegooijen2006B}
{Mackay}, D.~H., and A.~A. {van Ballegooijen}.
\newblock {Models of the Large-Scale Corona. II. Magnetic Connectivity and Open
  Flux Variation}.
\newblock \emph{\apj}, \textbf{642}, 1193--1204, 2006{\natexlab{b}}.
\newblock 10.1086/501043.

\bibitem[{{Mackay} et~al.(2016){Mackay}, {Yeates}, and {Bocquet}}]{Mackay2016}
{Mackay}, D.~H., A.~R. {Yeates}, and F.-X. {Bocquet}.
\newblock {Impact of an L5 Magnetograph on Nonpotential Solar Global Magnetic
  Field Modeling}.
\newblock \emph{\apj}, \textbf{825}, 131, 2016.
\newblock 10.3847/0004-637X/825/2/131.

\bibitem[{{Mays} et~al.(2015){Mays}, {Taktakishvili}, {Pulkkinen}, {MacNeice},
  {Rast{\"a}tter} et~al.}]{Mays2015}
{Mays}, M.~L., A.~{Taktakishvili}, A.~{Pulkkinen}, P.~J. {MacNeice},
  L.~{Rast{\"a}tter}, et~al.
\newblock {Ensemble Modeling of CMEs Using the WSA-ENLIL+Cone Model}.
\newblock \emph{\solphys}, \textbf{290}, 1775--1814, 2015.
\newblock 10.1007/s11207-015-0692-1, \eprint{1504.04402}.

\bibitem[{{Merkin} et~al.(2016){Merkin}, {Lionello}, {Lyon}, {Linker},
  {T{\"o}r{\"o}k}, and {Downs}}]{Merkin2016}
{Merkin}, V.~G., R.~{Lionello}, J.~G. {Lyon}, J.~{Linker}, T.~{T{\"o}r{\"o}k},
  and C.~{Downs}.
\newblock {Coupling of Coronal and Heliospheric Magnetohydrodynamic Models:
  Solution Comparisons and Verification}.
\newblock \emph{\apj}, \textbf{831}, 23, 2016.
\newblock 10.3847/0004-637X/831/1/23.

\bibitem[{{Michalek}(2006)}]{Michalek2006}
{Michalek}, G.
\newblock {An Asymmetric Cone Model for Halo Coronal Mass Ejections}.
\newblock \emph{\solphys}, \textbf{237}, 101--118, 2006.
\newblock 10.1007/s11207-006-0075-8, \eprint{0710.4537}.

\bibitem[{{Millward} et~al.(2013){Millward}, {Biesecker}, {Pizzo}, and
  {Koning}}]{Millward2013}
{Millward}, G., D.~{Biesecker}, V.~{Pizzo}, and C.~A. {Koning}.
\newblock {An operational software tool for the analysis of coronagraph images:
  Determining CME parameters for input into the WSA-Enlil heliospheric model}.
\newblock \emph{Space Weather}, \textbf{11}, 57--68, 2013.
\newblock 10.1002/swe.20024.

\bibitem[{{Na} et~al.(2017){Na}, {Moon}, and {Lee}}]{Na2017}
{Na}, H., Y.-J. {Moon}, and H.~{Lee}.
\newblock {Development of a Full Ice-cream Cone Model for Halo Coronal Mass
  Ejections}.
\newblock \emph{\apj}, \textbf{839}, 82, 2017.
\newblock 10.3847/1538-4357/aa697c.

\bibitem[{{Odstrcil}(2003)}]{Odstrcil2003}
{Odstrcil}, D.
\newblock {Modeling 3-D solar wind structure}.
\newblock \emph{Advances in Space Research}, \textbf{32}, 497--506, 2003.
\newblock 10.1016/S0273-1177(03)00332-6.

\bibitem[{{Odstrcil} et~al.(2004){Odstrcil}, {Riley}, and
  {Zhao}}]{Odstrcil2004}
{Odstrcil}, D., P.~{Riley}, and X.~P. {Zhao}.
\newblock {Numerical simulation of the 12 May 1997 interplanetary CME event}.
\newblock \emph{Journal of Geophysical Research (Space Physics)}, \textbf{109},
  A02116, 2004.
\newblock 10.1029/2003JA010135.

\bibitem[{{Odstr{\v c}il} and {Pizzo}(1999{\natexlab{a}})}]{OdstrcilPizzo1999a}
{Odstr{\v c}il}, D., and V.~J. {Pizzo}.
\newblock {Three-dimensional propagation of CMEs in a structured solar wind
  flow: 1. CME launched within the streamer belt}.
\newblock \emph{\jgr}, \textbf{104}, 483--492, 1999{\natexlab{a}}.
\newblock 10.1029/1998JA900019.

\bibitem[{{Odstr{\v c}il} and {Pizzo}(1999{\natexlab{b}})}]{OdstrcilPizzo1999b}
{Odstr{\v c}il}, D., and V.~J. {Pizzo}.
\newblock {Three-dimensional propagation of coronal mass ejections in a
  structured solar wind flow 2. CME launched adjacent to the streamer belt}.
\newblock \emph{\jgr}, \textbf{104}, 493--504, 1999{\natexlab{b}}.
\newblock 10.1029/1998JA900038.

\bibitem[{{Odstr{\v c}il} et~al.(1996){Odstr{\v c}il}, {Smith}, and
  {Dryer}}]{Odstrcil1996}
{Odstr{\v c}il}, D., Z.~{Smith}, and M.~{Dryer}.
\newblock {Distortion of the heliospheric plasma sheet by interplanetary
  shocks}.
\newblock \emph{\grl}, \textbf{23}, 2521--2524, 1996.
\newblock 10.1029/96GL00159.

\bibitem[{{Ouyang} et~al.(2015){Ouyang}, {Yang}, and {Chen}}]{Ouyang2015}
{Ouyang}, Y., K.~{Yang}, and P.~F. {Chen}.
\newblock {Is Flux Rope a Necessary Condition for the Progenitor of Coronal
  Mass Ejections?}
\newblock \emph{\apj}, \textbf{815}, 72, 2015.
\newblock 10.1088/0004-637X/815/1/72, \eprint{1511.01605}.

\bibitem[{{Pagano} et~al.(2013{\natexlab{a}}){Pagano}, {Mackay}, and
  {Poedts}}]{Pagano2013b}
{Pagano}, P., D.~H. {Mackay}, and S.~{Poedts}.
\newblock {Effect of gravitational stratification on the propagation of a CME}.
\newblock \emph{\aap}, \textbf{560}, A38, 2013{\natexlab{a}}.
\newblock 10.1051/0004-6361/201322036, \eprint{1310.6960}.

\bibitem[{{Pagano} et~al.(2013{\natexlab{b}}){Pagano}, {Mackay}, and
  {Poedts}}]{Pagano2013a}
{Pagano}, P., D.~H. {Mackay}, and S.~{Poedts}.
\newblock {Magnetohydrodynamic simulations of the ejection of a magnetic flux
  rope}.
\newblock \emph{\aap}, \textbf{554}, A77, 2013{\natexlab{b}}.
\newblock 10.1051/0004-6361/201220947.

\bibitem[{{Pagano} et~al.(2014){Pagano}, {Mackay}, and {Poedts}}]{Pagano2014}
{Pagano}, P., D.~H. {Mackay}, and S.~{Poedts}.
\newblock {Simulating AIA observations of a flux rope ejection}.
\newblock \emph{\aap}, \textbf{568}, A120, 2014.
\newblock 10.1051/0004-6361/201424019, \eprint{1407.8397}.

\bibitem[{{Pevtsov} et~al.(2016){Pevtsov}, {Bertello}, {MacNeice}, and
  {Petrie}}]{Pevtsov2016}
{Pevtsov}, A.~A., L.~{Bertello}, P.~{MacNeice}, and G.~{Petrie}.
\newblock {What if we had a magnetograph at Lagrangian L5?}
\newblock \emph{Space Weather}, \textbf{14}, 1026--1031, 2016.
\newblock 10.1002/2016SW001471.

\bibitem[{{Porth} et~al.(2014){Porth}, {Xia}, {Hendrix}, {Moschou}, and
  {Keppens}}]{Porth2014}
{Porth}, O., C.~{Xia}, T.~{Hendrix}, S.~P. {Moschou}, and R.~{Keppens}.
\newblock {MPI-AMRVAC for Solar and Astrophysics}.
\newblock \emph{\apjs}, \textbf{214}, 4, 2014.
\newblock 10.1088/0067-0049/214/1/4, \eprint{1407.2052}.

\bibitem[{{Rollett} et~al.(2016){Rollett}, {M{\"o}stl}, {Isavnin}, {Davies},
  {Kubicka}, {Amerstorfer}, and {Harrison}}]{Rollett2016}
{Rollett}, T., C.~{M{\"o}stl}, A.~{Isavnin}, J.~A. {Davies}, M.~{Kubicka},
  U.~V. {Amerstorfer}, and R.~A. {Harrison}.
\newblock {ElEvoHI: A Novel CME Prediction Tool for Heliospheric Imaging
  Combining an Elliptical Front with Drag-based Model Fitting}.
\newblock \emph{\apj}, \textbf{824}, 131, 2016.
\newblock 10.3847/0004-637X/824/2/131, \eprint{1605.00510}.

\bibitem[{{Sachdeva} et~al.(2015){Sachdeva}, {Subramanian}, {Colaninno}, and
  {Vourlidas}}]{Sachdeva2015}
{Sachdeva}, N., P.~{Subramanian}, R.~{Colaninno}, and A.~{Vourlidas}.
\newblock {CME Propagation: Where does Aerodynamic Drag 'Take Over'?}
\newblock \emph{\apj}, \textbf{809}, 158, 2015.
\newblock 10.1088/0004-637X/809/2/158, \eprint{1507.05199}.

\bibitem[{{Schrijver} et~al.(2015){Schrijver}, {Kauristie}, {Aylward},
  {Denardini}, {Gibson} et~al.}]{Schrijver2015}
{Schrijver}, C.~J., K.~{Kauristie}, A.~D. {Aylward}, C.~M. {Denardini}, S.~E.
  {Gibson}, et~al.
\newblock {Understanding space weather to shield society: A global road map for
  2015-2025 commissioned by COSPAR and ILWS}.
\newblock \emph{Advances in Space Research}, \textbf{55}, 2745--2807, 2015.
\newblock 10.1016/j.asr.2015.03.023, \eprint{1503.06135}.

\bibitem[{{Shiota} and {Kataoka}(2016)}]{ShiotaKataoka2016}
{Shiota}, D., and R.~{Kataoka}.
\newblock {Magnetohydrodynamic simulation of interplanetary propagation of
  multiple coronal mass ejections with internal magnetic flux rope
  (SUSANOO-CME)}.
\newblock \emph{Space Weather}, \textbf{14}, 56--75, 2016.
\newblock 10.1002/2015SW001308.

\bibitem[{{Tappin} and {Howard}(2009)}]{TappinHoward2009b}
{Tappin}, S.~J., and T.~A. {Howard}.
\newblock {Interplanetary Coronal Mass Ejections Observed in the Heliosphere:
  2. Model and Data Comparison}.
\newblock \emph{\ssr}, \textbf{147}, 55--87, 2009.
\newblock 10.1007/s11214-009-9550-5.

\bibitem[{{T{\"o}r{\"o}k} and {Kliem}(2005)}]{TorokKliem2005}
{T{\"o}r{\"o}k}, T., and B.~{Kliem}.
\newblock {Confined and Ejective Eruptions of Kink-unstable Flux Ropes}.
\newblock \emph{\apjl}, \textbf{630}, L97--L100, 2005.
\newblock 10.1086/462412, \eprint{astro-ph/0507662}.

\bibitem[{{T{\'o}th} et~al.(2005){T{\'o}th}, {Sokolov}, {Gombosi}, {Chesney},
  {Clauer} et~al.}]{Toth2005}
{T{\'o}th}, G., I.~V. {Sokolov}, T.~I. {Gombosi}, D.~R. {Chesney}, C.~R.
  {Clauer}, et~al.
\newblock {Space Weather Modeling Framework: A new tool for the space science
  community}.
\newblock \emph{Journal of Geophysical Research (Space Physics)}, \textbf{110},
  A12226, 2005.
\newblock 10.1029/2005JA011126.

\bibitem[{{T{\'o}th} et~al.(2012){T{\'o}th}, {van der Holst}, {Sokolov}, {De
  Zeeuw}, {Gombosi} et~al.}]{Toth2012}
{T{\'o}th}, G., B.~{van der Holst}, I.~V. {Sokolov}, D.~L. {De Zeeuw}, T.~I.
  {Gombosi}, et~al.
\newblock {Adaptive numerical algorithms in space weather modeling}.
\newblock \emph{Journal of Computational Physics}, \textbf{231}, 870--903,
  2012.
\newblock 10.1016/j.jcp.2011.02.006.

\bibitem[{{Tucker-Hood} et~al.(2015){Tucker-Hood}, {Scott}, {Owens}, {Jackson},
  {Barnard} et~al.}]{TuckerHood2015}
{Tucker-Hood}, K., C.~{Scott}, M.~{Owens}, D.~{Jackson}, L.~{Barnard}, et~al.
\newblock {Validation of a priori CME arrival predictions made using real-time
  heliospheric imager observations}.
\newblock \emph{Space Weather}, \textbf{13}, 35--48, 2015.
\newblock 10.1002/2014SW001106.

\bibitem[{{{\v Z}ic} et~al.(2015){{\v Z}ic}, {Vr{\v s}nak}, and
  {Temmer}}]{Zic2015}
{{\v Z}ic}, T., B.~{Vr{\v s}nak}, and M.~{Temmer}.
\newblock {Heliospheric Propagation of Coronal Mass Ejections: Drag-based Model
  Fitting}.
\newblock \emph{\apjs}, \textbf{218}, 32, 2015.
\newblock 10.1088/0067-0049/218/2/32, \eprint{1506.08582}.

\bibitem[{{Vourlidas} et~al.(2013){Vourlidas}, {Lynch}, {Howard}, and
  {Li}}]{Vourlidas2013}
{Vourlidas}, A., B.~J. {Lynch}, R.~A. {Howard}, and Y.~{Li}.
\newblock {How Many CMEs Have Flux Ropes? Deciphering the Signatures of Shocks,
  Flux Ropes, and Prominences in Coronagraph Observations of CMEs}.
\newblock \emph{\solphys}, \textbf{284}, 179--201, 2013.
\newblock 10.1007/s11207-012-0084-8, \eprint{1207.1599}.

\bibitem[{{Vr{\v s}nak} et~al.(2014){Vr{\v s}nak}, {Temmer}, {{\v Z}ic},
  {Taktakishvili}, {Dumbovi{\'c}}, {M{\"o}stl}, {Veronig}, {Mays}, and
  {Odstr{\v c}il}}]{Vr¨nak2014}
{Vr{\v s}nak}, B., M.~{Temmer}, T.~{{\v Z}ic}, A.~{Taktakishvili},
  M.~{Dumbovi{\'c}}, C.~{M{\"o}stl}, A.~M. {Veronig}, M.~L. {Mays}, and
  D.~{Odstr{\v c}il}.
\newblock {Heliospheric Propagation of Coronal Mass Ejections: Comparison of
  Numerical WSA-ENLIL+Cone Model and Analytical Drag-based Model}.
\newblock \emph{\apjs}, \textbf{213}, 21, 2014.
\newblock 10.1088/0067-0049/213/2/21.

\bibitem[{{Vr{\v s}nak} et~al.(2013){Vr{\v s}nak}, {{\v Z}ic}, {Vrbanec},
  {Temmer}, {Rollett} et~al.}]{Vr¨nak2013}
{Vr{\v s}nak}, B., T.~{{\v Z}ic}, D.~{Vrbanec}, M.~{Temmer}, T.~{Rollett},
  et~al.
\newblock {Propagation of Interplanetary Coronal Mass Ejections: The Drag-Based
  Model}.
\newblock \emph{\solphys}, \textbf{285}, 295--315, 2013.
\newblock 10.1007/s11207-012-0035-4.

\bibitem[{{Weinzierl} et~al.(2016{\natexlab{a}}){Weinzierl}, {Mackay},
  {Yeates}, and {Pevtsov}}]{Weinzierl2016b}
{Weinzierl}, M., D.~H. {Mackay}, A.~R. {Yeates}, and A.~A. {Pevtsov}.
\newblock {The Possible Impact of L5 Magnetograms on Non-potential Solar
  Coronal Magnetic Field Simulations}.
\newblock \emph{\apj}, \textbf{828}, 102, 2016{\natexlab{a}}.
\newblock 10.3847/0004-637X/828/2/102.

\bibitem[{{Weinzierl} et~al.(2016{\natexlab{b}}){Weinzierl}, {Yeates},
  {Mackay}, {Henney}, and {Arge}}]{Weinzierl2016a}
{Weinzierl}, M., A.~R. {Yeates}, D.~H. {Mackay}, C.~J. {Henney}, and C.~N.
  {Arge}.
\newblock {A New Technique for the Photospheric Driving of Non-potential Solar
  Coronal Magnetic Field Simulations}.
\newblock \emph{\apj}, \textbf{823}, 55, 2016{\natexlab{b}}.
\newblock 10.3847/0004-637X/823/1/55.

\bibitem[{{Xia} et~al.(2014){Xia}, {Keppens}, and {Guo}}]{Xia2014}
{Xia}, C., R.~{Keppens}, and Y.~{Guo}.
\newblock {Three-dimensional Prominence-hosting Magnetic Configurations:
  Creating a Helical Magnetic Flux Rope}.
\newblock \emph{\apj}, \textbf{780}, 130, 2014.
\newblock 10.1088/0004-637X/780/2/130, \eprint{1311.5478}.

\bibitem[{{Xie} et~al.(2004){Xie}, {Ofman}, and {Lawrence}}]{Xie2004}
{Xie}, H., L.~{Ofman}, and G.~{Lawrence}.
\newblock {Cone model for halo CMEs: Application to space weather forecasting}.
\newblock \emph{Journal of Geophysical Research (Space Physics)}, \textbf{109},
  A03109, 2004.
\newblock 10.1029/2003JA010226.

\bibitem[{{Xue} et~al.(2005){Xue}, {Wang}, and {Dou}}]{Xue2005}
{Xue}, X.~H., C.~B. {Wang}, and X.~K. {Dou}.
\newblock {An ice-cream cone model for coronal mass ejections}.
\newblock \emph{Journal of Geophysical Research (Space Physics)}, \textbf{110},
  A08103, 2005.
\newblock 10.1029/2004JA010698.

\bibitem[{{Yeates} et~al.(2010{\natexlab{a}}){Yeates}, {Attrill}, {Nandy},
  {Mackay}, {Martens}, and {van Ballegooijen}}]{Yeates2010}
{Yeates}, A.~R., G.~D.~R. {Attrill}, D.~{Nandy}, D.~H. {Mackay}, P.~C.~H.
  {Martens}, and A.~A. {van Ballegooijen}.
\newblock {Comparison of a Global Magnetic Evolution Model with Observations of
  Coronal Mass Ejections}.
\newblock \emph{\apj}, \textbf{709}, 1238--1248, 2010{\natexlab{a}}.
\newblock 10.1088/0004-637X/709/2/1238, \eprint{0912.3347}.

\bibitem[{{Yeates} and {Mackay}(2009{\natexlab{a}})}]{YeatesMackay2009}
{Yeates}, A.~R., and D.~H. {Mackay}.
\newblock {Initiation of Coronal Mass Ejections in a Global Evolution Model}.
\newblock \emph{\apj}, \textbf{699}, 1024--1037, 2009{\natexlab{a}}.
\newblock 10.1088/0004-637X/699/2/1024, \eprint{0904.4419}.

\bibitem[{{Yeates} and {Mackay}(2009{\natexlab{b}})}]{Yeates2009}
{Yeates}, A.~R., and D.~H. {Mackay}.
\newblock {Initiation of Coronal Mass Ejections in a Global Evolution Model}.
\newblock \emph{\apj}, \textbf{699}, 1024--1037, 2009{\natexlab{b}}.
\newblock 10.1088/0004-637X/699/2/1024, \eprint{0904.4419}.

\bibitem[{{Yeates} and {Mackay}(2012)}]{YeatesMackay2012}
{Yeates}, A.~R., and D.~H. {Mackay}.
\newblock {Chirality of High-latitude Filaments over Solar Cycle 23}.
\newblock \emph{\apjl}, \textbf{753}, L34, 2012.
\newblock 10.1088/2041-8205/753/2/L34, \eprint{1206.2327}.

\bibitem[{{Yeates} et~al.(2007){Yeates}, {Mackay}, and {van
  Ballegooijen}}]{Yeates2007}
{Yeates}, A.~R., D.~H. {Mackay}, and A.~A. {van Ballegooijen}.
\newblock {Modelling the Global Solar Corona: Filament Chirality Observations
  and Surface Simulations}.
\newblock \emph{\solphys}, \textbf{245}, 87--107, 2007.
\newblock 10.1007/s11207-007-9013-7, \eprint{0707.3256}.

\bibitem[{{Yeates} et~al.(2008){Yeates}, {Mackay}, and {van
  Ballegooijen}}]{Yeates2008}
{Yeates}, A.~R., D.~H. {Mackay}, and A.~A. {van Ballegooijen}.
\newblock {Modelling the Global Solar Corona II: Coronal Evolution and Filament
  Chirality Comparison}.
\newblock \emph{\solphys}, \textbf{247}, 103--121, 2008.
\newblock 10.1007/s11207-007-9097-0, \eprint{0711.2887}.

\bibitem[{{Yeates} et~al.(2010{\natexlab{b}}){Yeates}, {Mackay}, {van
  Ballegooijen}, and {Constable}}]{Yeates2010JGRA}
{Yeates}, A.~R., D.~H. {Mackay}, A.~A. {van Ballegooijen}, and J.~A.
  {Constable}.
\newblock {A nonpotential model for the Sun's open magnetic flux}.
\newblock \emph{Journal of Geophysical Research (Space Physics)}, \textbf{115},
  A09112, 2010{\natexlab{b}}.
\newblock 10.1029/2010JA015611, \eprint{1006.4011}.

\bibitem[{{Zhao} and {Dryer}(2014)}]{ZhaoDryer2014}
{Zhao}, X., and M.~{Dryer}.
\newblock {Current status of CME/shock arrival time prediction}.
\newblock \emph{Space Weather}, \textbf{12}, 448--469, 2014.
\newblock 10.1002/2014SW001060.

\bibitem[{{Zuccarello} et~al.(2015){Zuccarello}, {Aulanier}, and
  {Gilchrist}}]{Zuccarello2015}
{Zuccarello}, F.~P., G.~{Aulanier}, and S.~A. {Gilchrist}.
\newblock {Critical Decay Index at the Onset of Solar Eruptions}.
\newblock \emph{\apj}, \textbf{814}, 126, 2015.
\newblock 10.1088/0004-637X/814/2/126, \eprint{1510.03713}.

\end{thebibliography}


\end{document}